\documentclass[aps,prl,twocolumn,superscriptaddress,amssymb,amsfonts,longbibliography,10pt]{revtex4-2}

\usepackage{physics}
\usepackage{mathdots,bbm}
\usepackage{amsmath}
\usepackage{amssymb}
\usepackage{amsthm}
\usepackage[dvipsnames]{xcolor}
\usepackage{amsbsy}
\usepackage{amstext}
\usepackage[caption=false]{subfig}
\usepackage{graphicx}

\usepackage{color}
\usepackage{mathtools}
\usepackage{multirow}
\usepackage{amsfonts}
\usepackage{dcolumn}
\usepackage{bbold,bm}
\usepackage{tikz}
\usepackage[percent]{overpic}
\usepackage[scr=boondoxo]{mathalpha}
\usepackage[
    colorlinks,
    linkcolor={blue!80!black},
    citecolor={blue!80!black},
    urlcolor={blue!80!black}
]{hyperref}
\usepackage{xr-hyper}

\makeatletter
\newcommand{\hh}{\mathcal{H}}

\newcommand{\s}{\sigma}
\newcommand{\w}{\omega}

\newcommand{\G}{\gamma}

\newcommand{\bk}{{\bf k}}

\newcommand{\cP}{\hat{\mathcal P}}
\makeatother

\begin{document}
\allowdisplaybreaks

\title{Micromotion and topological constraints on Floquet-state quantum geometry}

\author{Arpit Raj}
\email{arpit.raj@uni-wuerzburg.de}
\affiliation{Institute for Theoretical Physics and Astrophysics, University of W\"urzburg, D-97074 W\"urzburg, Germany}
\affiliation{W\"urzburg-Dresden Cluster of Excellence ctd.qmat, D-97074 W\"urzburg, Germany}

\author{Johannes Mitscherling}
\email{johannes.mitscherling@uni-wuerzburg.de}
\affiliation{Institute for Theoretical Physics and Astrophysics, University of W\"urzburg, D-97074 W\"urzburg, Germany}
\affiliation{W\"urzburg-Dresden Cluster of Excellence ctd.qmat, D-97074 W\"urzburg, Germany}
\affiliation{Max Planck Institute for the Physics of Complex Systems, N\"othnitzer Str. 38, 01187 Dresden, Germany}

\author{Bj\"orn Trauzettel}
\email{bjoern.trauzettel@uni-wuerzburg.de}
\affiliation{Institute for Theoretical Physics and Astrophysics, University of W\"urzburg, D-97074 W\"urzburg, Germany}
\affiliation{W\"urzburg-Dresden Cluster of Excellence ctd.qmat, D-97074 W\"urzburg, Germany}

\begin{abstract}
The quantum geometry of Floquet-Bloch states in periodically driven crystals evolves throughout each drive cycle. We show that Floquet topology and micromotion constrain this evolution. Using a projector formulation that separates stroboscopic and micromotion contributions, we bound the deviation of the integrated quantum metric of any Floquet-band subspace from its stroboscopic value and derive a geometric speed limit on its evolution. In chiral Floquet systems, these bounds extend the topological lower bounds set by the $0$- and $\pi$-gap invariants at two symmetry-selected times into finite intervals of the drive, a dynamical topological obstruction. When both invariants are nonzero, it can force a minimum intra-period change in the metric, driven by micromotion. Because the integrated metric is a periodic lower bound on the total spread of time-evolved Wannier functions, such a change can make this spread breathe. An exactly solvable driven spin chain realizes this obstruction and further hosts a flat-band point where each band's metric equals the breathing of its compact localized state. This equality provides a direct geometric characterization of local micromotion, accessible through local spin correlators.
\end{abstract}

\maketitle

\textit{Introduction.---} Periodically driven lattice systems exhibit topology with no static counterpart~\cite{harper2020topology}. A minimal chiral-symmetric Floquet chain already hosts anomalous $\pi$-modes at the edge of the Floquet Brillouin zone (BZ), characterized by an integer $\nu_\pi$ in addition to the invariant $\nu_0$ of the conventional gap~\cite{asboth2014chiral}. Further Floquet invariants arise in higher dimensions~\cite{kitagawa2010topological, rudner2013anomalous, wintersperger2020realization}. In static crystals, the quantum metric~\cite{provost1980riemannian} plays a dual role: its momentum integral bounds topological invariants~\cite{roy2014band, ozawa2021relations, Mera2022, onishi2024fundamental, mera2022relating} and equals the gauge-invariant spread of Wannier functions, i.e., the variance of the polarization distribution of insulators~\cite{kohn1964theory, resta1999electron, souza2000polarization, resta2011, marzari2012maximally}. Floquet theory itself has an inherent geometric structure~\cite{schindler2025geometric}. Within each drive cycle, the Floquet-Bloch states and their quantum geometry evolve through the micromotion, which calls for a geometric description beyond stroboscopic properties.

Recent work has extended quantum-geometric concepts to the time domain. The micromotion-operator metric has a quantum volume bounded from below by Floquet invariants in certain symmetry classes~\cite{he2026floquet}, and the momentum-time Berry connection of Floquet states defines a Chern-Simons axion coupling quantized by the winding of the micromotion operator in anomalous two-dimensional phases~\cite{gavensky2026quantized}. Time-dependent quantum geometric tensors have been defined for the instantaneous response of insulators~\cite{verma2025instantaneous} and for Floquet-Bloch states~\cite{dabiri2026time}, and the geometry of time-dependent density matrices extends them to many-body systems~\cite{guan2026exploring}. Projector-based formulations give access to geometric quantities beyond the quantum geometric tensor~\cite{Avdoshkin2025, Mitscherling2025}. These works characterize several aspects of Floquet-state geometry and its topological content but leave open two questions: (i) How far and how fast can the Floquet-state geometry change within a drive cycle, and (ii) how does Floquet topology constrain it beyond stroboscopic times?
\begin{figure}[t!]
    \centering
    \includegraphics[width=0.99\linewidth]{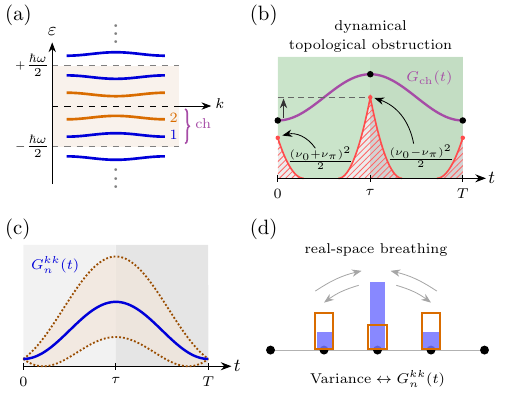}
    \caption{Projectors onto Floquet bands and band subspaces define a quantum metric that characterizes the time dynamics of Floquet-Bloch states (a). Through micromotion, the gap invariants bound the chiral-subspace metric $G_{\rm ch}(t)$ from below in finite time windows around the symmetric times $0$ and $\tau$, a dynamical topological obstruction that can enforce time variation of quantum geometry throughout the drive cycle (b). The band-resolved metric is bounded by the micromotion geometry (c) and, in a flat band, equals the breathing variance of a compact localized state (d).}
    \label{fig:main}
\end{figure}

In this Letter, we present a unifying quantum geometric theory for the entire drive cycle, establishing bounds on the evolution of Floquet-state geometry arising from micromotion-operator geometry and Floquet topology. To this end, we express the time-evolved Floquet projector in terms of the micromotion operator in a stroboscopic frame, thereby separating the quantum metric of the Floquet subspace of interest into stroboscopic, micromotion, and interference terms. Bounding the interference term constrains the integrated metric's deviation from its stroboscopic value, while the rate of micromotion-induced position displacement sets a geometric speed limit on its evolution. In chiral chains, the two symmetry-selected Floquet frames can carry different windings, yet their metrics lie on the same micromotion curve. Combined with the micromotion bound, their topological bounds constrain the integrated metric over finite time intervals, suggesting a dynamical topological obstruction, as sketched in Fig.~\ref{fig:main}. An exact variance decomposition connects the time dynamics of quantum geometry to the spread of time-evolved Wannier functions. We exemplify these relations in a solvable driven chiral spin chain, including a flat-band point where the integrated metric exactly tracks the breathing of compact localized states.

\smallskip
\textit{Floquet quantum geometry.---} We consider the time-periodic Bloch Hamiltonian $\hat H(\bk,t)=\hat H(\bk,t+T)$ and its evolution operator $\hat U(\bk,t_2,t_1)=\mathcal{T} e^{-\frac{i}{\hbar}\int_{t_1}^{t_2}\dd{t'} \hat H(\bk,t')}$, where $\mathcal{T}$ denotes time ordering. At reference time $t_0$, the one-period evolution operator has the spectral decomposition $\hat U(\bk,t_0+T,t_0)=\sum_n e^{-i\varepsilon_n(\bk)T/\hbar} \hat P_n(\bk,t_0)$, where, for nondegenerate Floquet bands, $\hat P_n(\bk,t_0)=|\varphi_n(\bk,t_0)\rangle\langle\varphi_n(\bk,t_0)|$ projects onto a time-periodic Floquet state satisfying $(\hat H(\bk,t)-i\hbar \, \partial_t)|\varphi_n(\bk,t)\rangle =\varepsilon_n(\bk)|\varphi_n(\bk,t)\rangle$~\cite{shirley1965solution, sambe1973steady, Grifoni1998}. Defining the corresponding Floquet Hamiltonian $\hat H_F(\bk,t_0)=\frac{i\hbar}{T}\ln \hat U(\bk,t_0+T,t_0)$, whose branch cut is fixed in End Matter~\ref{em:branchcut}, we separate the time-evolution operator as
\begin{align}
    \hat U(\bk,t,t_0)=\hat V(\bk,t,t_0)\,e^{-\frac{i}{\hbar}\hat H_F(\bk,t_0)\,(t-t_0)} \, ,
    \label{eqn:micromotionOperatorDefinition}
\end{align}
where the Floquet micromotion operator $\hat V(\bk,t,t_0)$ satisfies $\hat V(\bk,t_0,t_0) = \mathbbm{1}$ and $\hat V(\bk,t+T,t_0) = \hat V(\bk,t,t_0)$~\cite{goldman2014periodically, bukov2015universal, eckardt2015high}.

The Floquet projector is a density matrix onto a pure state and, thus, evolves in time through the quantum Liouville equation $i\hbar\,\partial_t \cP_n(\bk,t,t_0) = \big[\hat H(\bk,t),\cP_n(\bk,t,t_0)\big]$ with initial condition $\cP_n(\bk,t_0,t_0) = \hat P_n(\bk,t_0)$. The time-evolved Floquet projector takes the form 
\begin{align}
    \cP_{n}(\bk,t,t_0)= \hat V(\bk,t,t_0)\,\hat P_n(\bk,t_0)\,\hat V^\dagger(\bk,t,t_0) \, ,
    \label{eqn:timeFloquetProjector}
\end{align}
which follows from Eq.~\eqref{eqn:micromotionOperatorDefinition} and the solution $\cP_n=\hat U\hat P_n\hat U^\dagger$ of the Liouville equation, because $\hat H_F(\bk,t_0)$ commutes with $\hat P_n(\bk,t_0)$. Since $\hat U(\bk,t+T,t) = \hat U(\bk,t,t_0)\,\hat U(\bk,t_0+T,t_0)\,\hat U^\dag(\bk,t,t_0)$, the evolved projector $\cP_n(\bk,t,t_0)$ is the spectral projector onto quasienergy $\varepsilon_n(\bk)$ of the one-period evolution operator based at time $t$, and is therefore independent of the reference time $t_0$. We retain $t_0$ to specify the stroboscopic frame in which we decompose its geometry [Eq.~\eqref{eq:g_decomposition}].

The projector $\cP_n(\bk,t,t_0)$ is invariant under $\hat V\to\hat V\hat{\mathcal{M}}$ for any unitary $\hat{\mathcal{M}}$ commuting with $\hat P_{n}(\bk,t_0)$, so evolving with $\hat V$, $\hat U$, or the Wilson-line operator of Ref.~\cite{schindler2025geometric} yields the same time-evolved Floquet projector and corresponding quantum geometry. We use $\hat V$ because it makes each term in the metric decomposition [Eq.~\eqref{eq:g_decomposition}] $T$-periodic.

The quantum geometric tensor associated with the time-evolved Floquet projector reads
\begin{align}
\begin{split}
    Q^{\alpha\beta}_{n}(\bk,t,t_0) &= \text{Tr}\big[\cP_n(\bk,t,t_0)\,\partial_\alpha\cP_n(\bk,t,t_0) \\
    &\hspace{3cm} \times \partial_\beta\cP_n(\bk,t,t_0)\big] ,
\end{split}
\end{align}
where Greek indices $\alpha,\beta\in\{k_1,\ldots,k_d,t\}$ denote momentum and time coordinates, while Latin indices $a,b$ label momentum directions only. Its real and imaginary parts define the Floquet-state quantum metric and Berry curvature, respectively, through $Q^{\alpha\beta}_{n}=g^{\alpha\beta}_{n}-\frac{i}{2}\Omega^{\alpha\beta}_{n}$. For a single band, this is the time-dependent quantum geometric tensor of Floquet-Bloch states~\cite{dabiri2026time}.

The same construction generalizes to any multiband projector $\hat{\mathcal{P}}_S=\sum_{n\in S}\hat{\mathcal{P}}_n$ over a set of Floquet bands as well as to geometric quantities beyond the quantum metric and Berry curvature such as higher-order and multi-state geometric invariants relevant, e.g., for optical responses~\cite{Avdoshkin2025, Mitscherling2025}. At $t=t_0$, these geometric quantities reduce to their
stroboscopic counterparts.

To connect the Floquet-state quantum metric to real-space observables, we define its BZ integral
\begin{align}
    G^{ab}_{n}(t,t_0) = \frac{V_{\rm uc}}{(2\pi)^d} \int_{\rm BZ} \dd[d]{\bk} \, g^{ab}_{n}(\bk,t,t_0) \, ,
    \label{eq:Gint}
\end{align}
where $V_{\rm uc}$ is the unit-cell volume. This integral is the gauge-invariant part of the position covariance of the Wannier functions built from the evolved Floquet-Bloch states $\ket{\chi_n(\bk,t,t_0)}=\hat V(\bk,t,t_0)\ket{\varphi_n(\bk,t_0)}$, whose projector is $\cP_n(\bk,t,t_0)=\ket{\chi_n(\bk,t,t_0)}\bra{\chi_n(\bk,t,t_0)}$~\cite{marzari1997, resta2011, marzari2012maximally}. For an isolated subspace $S$, we define $g^{ab}_S=\frac{1}{2}\Tr\big[\partial_a\cP_S\,\partial_b\cP_S\big]$ and $G^{ab}_S$ analogously.

\smallskip
\textit{Relation to micromotion-operator geometry.---} The projector representation in Eq.~\eqref{eqn:timeFloquetProjector} allows us to
separate the stroboscopic and micromotion contributions to the
quantum metric. For its momentum-diagonal components, we obtain
\begin{align}
    & g^{aa}_n(\bk,t,t_0) = g^{aa}_n(\bk,t_0) \nonumber \\
    &\quad +\text{Tr}\Big[\hat P_n(\bk,t_0)\,\hat\xi^a_V(\bk,t,t_0)\,{\hat{\bar{P}}}_n(\bk,t_0)\,\hat \xi^a_V(\bk,t,t_0)\Big] \nonumber \\
    &\quad +\text{Tr}\Big[\hat {\mathcal{L}}^a_n(\bk,t_0)\,{\hat \xi}^a_V(\bk,t,t_0)\Big] \, ,
    \label{eq:g_decomposition}
\end{align}
where $\hat{\bar{P}}_n\equiv \mathbbm{1}-\hat P_n$ and $g^{aa}_{n}(\bk,t_0)\equiv g^{aa}_{n}(\bk,t_0,t_0)$. See End Matter~\ref{em:Gdecomposition} for the derivation and the Supplemental Material (SM) for all momentum components.
The first term is the stroboscopic metric, $g^{ab}_n(\bk,t_0) = \frac{1}{2}\Tr\big[\partial_a\hat P_n(\bk,t_0)\,\partial_b\hat P_n(\bk,t_0)\big]$. 
Micromotion enters through the Maurer--Cartan form $\hat\xi^a_V(\bk,t,t_0) = i\,\hat V^\dagger(\bk,t,t_0)\,\partial_a\hat V(\bk,t,t_0)$~\cite{Ahn2022}, which we call the \emph{micromotion connection}. In the site-resolved Bloch representation (End Matter~\ref{em:branchcut}), where $\hat X_a$ acts as $i\partial_{k_a}$, it captures the shift of the position operator under micromotion: $\hat V^\dagger\hat X_a\hat V=\hat X_a+\hat\xi^a_V$. Its interband weight is the second term in Eq.~\eqref{eq:g_decomposition}. The micromotion connection is also the building block of the micromotion-operator metric~\cite{he2026floquet}, $\frac{1}{2}\Tr\big[\hat \xi^a_V\hat \xi^a_V\big] = \frac{1}{2}\Tr\big[\partial_a\hat V^\dagger\,\partial_a\hat V\big]$.
The third term in Eq.~\eqref{eq:g_decomposition} describes the interference between the stroboscopic and micromotion geometries through the overlap of $\hat\xi^a_V$ with the band-resolved adiabatic connection operator of the Floquet projector, $\hat{\mathcal{L}}^a_n(\bk,t_0) = -i\big[\hat P_n(\bk,t_0),\partial_a\hat P_n(\bk,t_0)\big]$~\cite{Mitscherling2025Orbital}.

\smallskip
\textit{Micromotion bounds on metric evolution.---} The decomposition in Eq.~\eqref{eq:g_decomposition} bounds how far the integrated metric can depart from its stroboscopic value $G^{aa}_n(t_0)\equiv G^{aa}_n(t_0,t_0)$. We denote by $M^{aa}_n(t,t_0)\geq 0$ the BZ integral of the second term in Eq.~\eqref{eq:g_decomposition}, and by $G^{aa}_V(t,t_0)$ the BZ integral of $\frac{1}{2}\Tr\big[\partial_a\hat V^\dagger\,\partial_a\hat V\big]$, the micromotion-operator metric of Ref.~\cite{he2026floquet}, both normalized as in Eq.~\eqref{eq:Gint}. Bounding the interference term by two Cauchy--Schwarz inequalities (End Matter~\ref{em:Mbound}) gives
\begin{align}
    \Big|\sqrt{G^{aa}_n(t,t_0)}-\sqrt{G^{aa}_n(t_0)}\Big| \leq \sqrt{M^{aa}_n(t,t_0)} \leq \sqrt{G^{aa}_V(t,t_0)} ,
    \label{eq:micromotion-bound}
\end{align}
for the micromotion operator of any branch cut whose gap is open (End Matter~\ref{em:branchcut}). Its two-sided form also bounds $G^{aa}_n(t,t_0)$ from below when $M^{aa}_n(t,t_0)>G^{aa}_n(t_0)$, see Eq.~\eqref{eq:envelope} in End Matter~\ref{em:Mbound}. We point out that the second bound in Eq.~\eqref{eq:micromotion-bound} is weaker but band independent. 

Since $\cP_n(\bk,t,t_0)$ obeys a Liouville equation, the rate of change of the integrated metric obeys a geometric speed limit (End Matter~\ref{em:speedbound}),
\begin{align}
    \left| \dv{t}\sqrt{G^{aa}_n(t,t_0)}\right| \leq \sqrt{K^{aa}_n(t,t_0)} \, , \label{eq:speed-limit}
\end{align}
where $K^{aa}_n(t,t_0)$ is the integral of $\text{Tr}\Big[\hat P_n\partial_t\hat\xi^a_V{\hat{\bar{P}}}_n\partial_t\hat \xi^a_V\Big]$, the interband weight of $\partial_t \hat\xi^a_V$, normalized as in Eq.~\eqref{eq:Gint}.

Bounding the trace by the operator norm gives $|\dv*{\sqrt{G^{aa}_n}}{t}|\leq v^a_{\max}=\max_{\bk,t}\|\partial_t\hat\xi^a_V\|_{\rm op}$, the maximal velocity of the micromotion-induced displacement $\hat V^\dagger\hat X_a\hat V-\hat X_a$. Equations~\eqref{eq:g_decomposition}--\eqref{eq:speed-limit} extend to any isolated subspace $S$ of $N_S$ Floquet bands upon replacing $n\to S$ and, in the velocity bound, $v^a_{\max}\to\sqrt{N_S}\,v^a_{\max}$.

\smallskip
\textit{Dynamical topological obstruction.---} In a chiral-symmetric driven system, the evolution operators $U_1$ over $(t_*,t_*+\tau)$ and $U_2$ over $(t_*+\tau,t_*+T)$ satisfy $U_2=\Gamma U_1^\dagger\Gamma$ for suitable $t_*$ and $0<\tau<T$, with $\Gamma$ the chiral operator~\cite{asboth2014chiral, asboth2013bulk}. Since the reference time of the Floquet projector is arbitrary, we set $t_0=t_*$ and call $t_0$ and $t_0+\tau$ the \emph{symmetric times}. The one-period evolution operators starting at them, $U\equiv U(\bk,t_0+T,t_0)=U_2U_1$ and $\widetilde U\equiv U(\bk,t_0+\tau+T,t_0+\tau)=U_1U_2=U_1UU_1^\dagger$, define the \emph{symmetric-time frames}. They share the quasienergy spectrum, but the $\bk$-dependent conjugation generically changes their eigenstates~\cite{xu2022topological}; in one dimension, their effective Hamiltonians carry windings $\nu_0+\nu_\pi$ and $\nu_0-\nu_\pi$, respectively~\cite{asboth2014chiral}.

Both frames obey $\Gamma U\Gamma=U^\dagger$ and $\Gamma\widetilde U\Gamma=\widetilde U^\dagger$, which pairs the quasienergies as $\pm\varepsilon$. With both the $0$- and $\pi$-gaps open, the spectrum thus splits into two subspaces of dimension $N/2$, with $N$ the number of bands. We define the \emph{chiral projector} $\hat P_{\rm ch}=\sum_{\varepsilon_n<0}\hat P_n$ onto the negative-quasienergy bands of the zone $(-\hbar\w/2,\hbar\w/2]$ (see End Matter~\ref{em:branchcut}). It satisfies $\Gamma\hat P_{\rm ch}\Gamma=\mathbbm{1}-\hat P_{\rm ch}$, so $\Gamma$ exchanges the two halves. Since complementary subspaces always have the same metric, $G_{\rm ch}(t)\equiv G^{kk}_{\rm ch}(t,t_0)$ characterizes the bipartition itself.

The chiral projector of $\widetilde U$ is the evolved projector $\cP_{\rm ch}(\bk,t_0+\tau,t_0)$, so the metrics of the two frames, $G_{\rm ch}(t_0)$ and $G_{\rm ch}(t_0+\tau)$, are two points on the same micromotion curve. A generic time frame admits no splitting into $\Gamma$-conjugate halves (End Matter~\ref{em:chiralspinchain}), so the direct metric-winding bounds arise only at the symmetric times.

Applying the Cauchy--Schwarz inequality to the winding number of each frame yields
\begin{align}
    \begin{split}
    G_{\rm ch}(t_0)&\ge \frac{V_{\rm uc}^2}{2N}\,(\nu_0+\nu_\pi)^2 \,,\\
    G_{\rm ch}(t_0+\tau)&\ge \frac{V_{\rm uc}^2}{2N}\,(\nu_0-\nu_\pi)^2 \,,
    \end{split}
    \label{eq:two-bounds}
\end{align}
see End Matter~\ref{em:chiralTopoBound}. Saturation requires the chiral connection of the respective frame to be a $k$-independent multiple of the identity. Furthermore, since both invariants are integers, the metric at a symmetric time bounds its frame winding by $\bar\nu(t_s)=\big\lfloor\sqrt{2N\,G_{\rm ch}(t_s)}/V_{\rm uc}\big\rfloor$, which confines $|\nu_0|,|\nu_\pi|\leq \big\lfloor[\bar\nu(t_0)+\bar\nu(t_0+\tau)]/2\big\rfloor$. The metric $G_{\rm ch}$ is experimentally accessible~\cite{ozawa2018extracting, tan2019experimental, gianfrate2020measurement, yu2024experimental}.

Although Eq.~\eqref{eq:two-bounds} appears as two static bounds, its dynamical content follows in combination with Eq.~\eqref{eq:micromotion-bound}. Taking a symmetric time $t_s$ as the reference, it yields
\begin{align}
    \frac{V_{\rm uc}}{\sqrt{2N}}\,|\nu_s| \leq \sqrt{G_{\rm ch}(t)} + \sqrt{M^{kk}_{\rm ch}(t,t_s)} \, ,
    \label{eq:shared}
\end{align}
where $\nu_s=\nu_0\pm\nu_\pi$ is the winding at the chosen symmetric time $t_s$. The topology fixed at $t_s$ thus bounds the sum of the state metric and the micromotion weight at every instant of time, so the two cannot both be arbitrarily small. Since $M^{kk}_{\rm ch}(t_s,t_s)=0$, Eq.~\eqref{eq:shared} reduces to Eq.~\eqref{eq:two-bounds} at $t_s$ and bounds $G_{\rm ch}$ from below in finite windows around it, which constitutes a form of \emph{dynamical topological obstruction} [Fig.~\ref{fig:main}(b)]. Unlike a Chern number in two dimensions, which is carried by the evolved projector at every instant of time~\cite{dabiri2026time}, the chiral gap invariants are not instantaneous properties of the projector. Equation~\eqref{eq:shared} nevertheless carries them into the micromotion.

The two bounds in Eq.~\eqref{eq:two-bounds} differ by $2V_{\rm uc}^2|\nu_0\nu_\pi|/N$ and thus coincide unless both gaps are topological, a situation without static counterpart. The larger bound is then $V_{\rm uc}^2(|\nu_0|+|\nu_\pi|)^2/(2N)$. Whenever $G_{\rm ch}$ at the symmetric time with the smaller bound lies below this value, the metric must change within each drive cycle, and by Eq.~\eqref{eq:micromotion-bound}, $\sqrt{M^{kk}_{\rm ch}}$ must be at least as large as the change in $\sqrt{G_{\rm ch}}$.

\begin{figure}[t]
    \centering
    \includegraphics[width=0.99\linewidth]{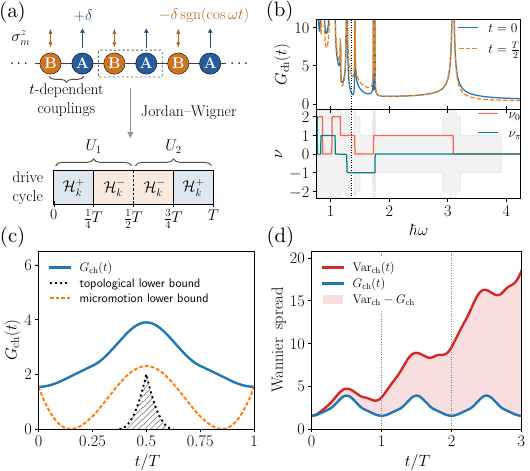}
    \caption{(a)~Driven spin chain~\eqref{eq:spinHam0} with a two-site unit cell, static field on $A$ (blue), and square-wave field on $B$ (orange); $U_1$ and $U_2$ denote half-period evolutions. (b)~$G_{\rm ch}(0)$ (solid) and $G_{\rm ch}(T/2)$ (dashed) versus $\w$ at $\G=\delta=1$ (upper), with gap invariants $\nu_0,\nu_\pi$ (lower). The shaded region is allowed by Eq.~\eqref{eq:two-bounds}, the dotted line marks $\hbar\w=1.35$. (c)~$G_{\rm ch}(t)$ at $\hbar\w=1.35$, with the micromotion lower bound~\eqref{eq:envelope} (dashed, $t_0=0$) and topological lower bound~\eqref{eq:shared} (dotted, $t_s=T/2$). Shading marks the excluded region. (d)~Total variance $\mathrm{Var}_{\rm ch}(t)$ of the two chiral MLWFs associated with the unit cell at sites 49 and 50, constructed at $t=0$, compared with $G_{\rm ch}(t)$.}
    \label{fig:4B_results_strobo}
\end{figure}

\smallskip
\textit{Driven chiral spin chain.---} An exactly solvable realization of both $0$- and $\pi$-gap topology and flat bands hosting compact localized states (CLS) is the periodically driven spin-$\tfrac12$ chain
\begin{align}
 H(t) =& \sum_m\bigg[ \frac{1-\G\sin{\w t}}{4}\,\s^x_m\s^x_{m+1}
      + \frac{1+\G\sin{\w t}}{4}\,\s^y_m\s^y_{m+1} \nonumber \\
      &- \frac{\G\cos{\w t}}{4}\,(\s^x_{m+1}\s^y_m+\s^x_m\s^y_{m+1})
      - \frac{\delta_m(t)}{2}\,\s^z_m \bigg],
\label{eq:spinHam0}
\end{align}
with a sublattice-asymmetric field $\delta_m(t)=\delta$ on even sites (A) and $-\delta\,\mathrm{sgn}(\cos\w t)$ on odd sites (B) [Fig.~\ref{fig:4B_results_strobo}(a)]. The bond terms combine into isotropic $\rm XY$ exchange and a uniaxial anisotropy $\G$ whose axis precesses in the $xy$ plane at $\w/2$. Since all couplings act between nearest neighbors, a Jordan--Wigner transformation maps the chain exactly onto a driven $p$-wave chain of free fermions. Rotating to the frame of the precessing axis, in which we quote all invariants, makes the bond terms static and adds $\hbar\w/2$ on-site offsets, so that only the longitudinal field remains driven. This yields the piecewise-constant $4\times4$ BdG Bloch Hamiltonian
\begin{align}
    \hh_k^{\pm}=\mqty(
        \delta-\tfrac{\hbar\w}{2} & \cos k & 0 & \G\sin k\\
        \cos k & \mp\delta-\tfrac{\hbar\w}{2} & \G\sin k & 0\\
        0 & \G\sin k & -\delta+\tfrac{\hbar\w}{2} & -\cos k\\
        \G\sin k & 0 & -\cos k & \pm\delta+\tfrac{\hbar\w}{2}),
        \label{eq:bloch_ham}
\end{align}
with $\hh_k^{+}$ on $0<t<\tfrac{T}{4}$ and $\tfrac{3T}{4}<t<T$, $\hh_k^{-}$ on $\tfrac{T}{4}<t<\tfrac{3T}{4}$, and $k\in[-\tfrac\pi2,\tfrac\pi2)$ since $V_{\rm uc}=2$ in units of the site spacing. The model is chiral-symmetric, with symmetric times $t_0=0$ and $t_0+\tau=T/2$, and lies in class BDI (see End Matter~\ref{em:chiralspinchain}), which does not further constrain our analysis since the bounds [Eqs.~\eqref{eq:two-bounds} and \eqref{eq:shared}] require only chiral symmetry. The sublattice-asymmetric drive is essential: for a static field, the two symmetric-time frames coincide, which forces $\nu_0\nu_\pi=0$, and for a uniform static field the model reduces to the two-band model of Ref.~\cite{yang2019floquet} (SM). Four bands moreover allow band-resolved dynamics within the chiral subspace, which is absent for $N=2$, where $G^{kk}_n=G_{\rm ch}$.

Figure~\ref{fig:4B_results_strobo}(b) shows $G_{\rm ch}$ at both symmetric times together with the windings. $G_{\rm ch}$ diverges wherever a quasienergy gap closes and a winding jumps~\cite{longwen2024quantum}, and is thus sensitive to both gaps. At $\hbar\w=1.35$, where $\nu_0=-\nu_\pi=1$, Eq.~\eqref{eq:two-bounds} imposes no nonzero lower bound at $t=0$, where $G_{\rm ch}(0)=1.54$, but requires $G_{\rm ch}(T/2)\geq 2$. Topology therefore enforces a metric increase of at least $0.46$ between $t=0$ and $T/2$. The metric indeed reaches $3.9$ at $T/2$. In Fig.~\ref{fig:4B_results_strobo}(c), the topological lower bound of Eq.~\eqref{eq:shared} is nonzero for $0.375\leq t/T\leq0.625$, and the micromotion lower bound from $t_0=0$ reaches $2.30$ at $T/2$.

\smallskip
\textit{Tracking micromotion through real-space breathing of Floquet Wannier states.---} The Floquet metric acquires a real-space meaning through Wannier functions. For an isolated one-dimensional band $n$, a Wannier function constructed at $t_0$ from $\ket{\varphi_n}\equiv\ket{\varphi_n(k,t_0)}$ and evolved with $\hat U$ has the position variance (End Matter~\ref{em:wannierGt})
\begin{align}
    \mathrm{Var}_n[x](t)=G^{kk}_n(t,t_0)+\mathrm{Var}_k[\mathcal A_n(k,t)] ,
    \label{eq:VarSplit}
\end{align}
where $\mathrm{Var}_k$ is the BZ variance and $\mathcal A_n(k,t)$ the Berry connection of the evolved Bloch states,
\begin{align}
    \mathcal A_n(k,t) = \mathcal A_n(k,t_0)+\mel{\varphi_n}{\hat\xi^k_V}{\varphi_n} + \frac{t-t_0}{\hbar}\,\partial_k\varepsilon_n(k),
\label{eq:Aevolved}
\end{align}
with $\mathcal A_n(k,t_0)=i\braket{\varphi_n}{\partial_k \varphi_n}$. The three terms correspond to the initial Berry connection, the band-diagonal part of the micromotion connection which displaces and reshapes the Wannier function, and the group velocity, respectively.
The metric is thus the periodic lower bound on the spread, which extends the static metric-localization link~\cite{resta2011, ozawa2019probing, oliveira2025realspace} to driven crystals and makes precise its interpretation as a polarization fluctuation~\cite{dabiri2026time}. For dispersive bands, the group-velocity term produces ballistic spreading, quadratic in the number of elapsed cycles at fixed drive phase (End Matter~\ref{em:wannierGt}), which is absent for flat bands.

For an isolated subspace $S$, the total spread of any $N_S$ Wannier functions spanning $S$ is bounded from below by $G^{kk}_S(t,t_0)$, with equality at $t_0$ for maximally localized Wannier functions in one dimension~\cite{marzari1997}. Thus, for a basis maximally localized at $t_0$, a topologically enforced increase of $G^{kk}_S$ necessarily increases the total Wannier spread [Fig.~\ref{fig:4B_results_strobo}(d)]. If the bands in $S$ are individually isolated, Eq.~\eqref{eq:VarSplit} gives for their band Wannier functions
\begin{align}
    \max_{n\in S}\mathrm{Var}_n[x](t) \geq\frac{1}{N_S}\sum_{n\in S}G^{kk}_n(t,t_0) \geq\frac{G^{kk}_S(t,t_0)}{N_S},
    \label{eq:Var-chain}
\end{align}
where the last inequality holds because the band metrics, unlike the subspace metric, also include nonnegative contributions from the interband Berry connection within $S$. Thus, at least one band Wannier function has an rms width of at least $\sqrt{G^{kk}_S/N_S}$. At either symmetric time, Eq.~\eqref{eq:two-bounds} therefore requires
\begin{align}
\max_{n\in \rm ch}\sqrt{\mathrm{Var}_n[x](t_s)} \geq \frac{V_{\rm uc}|\nu_s|}{N}.
\label{eq:width-bound}
\end{align}
For compactly supported Wannier functions, this implies a minimum support length (End Matter~\ref{em:wannierGt}).

\begin{figure}[t]
    \centering
    \includegraphics[width=0.99\linewidth]{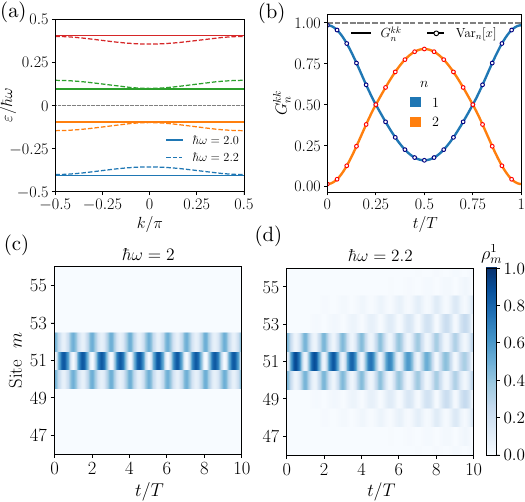}
    \caption{(a) Flat and nearly flat quasienergy spectrum. (b) For a flat band, $G^{kk}_n(t)$ (lines) coincides with the real-space variance $\mathrm{Var}_n[x](t)$ (dots). The dashed gray line denotes $G_{\rm ch}(t)$. (c) Periodic breathing of the site-resolved density of a band-1 CLS initially localized on sites 50--52 at $\hbar\omega=2$. (d) Spreading of a band-1 MLWF initially centered at site 51 at $\hbar\omega=2.2$. The chain contains 100 sites.}
    \label{fig:4B_results_micro}
\end{figure}

The chain realizes breathing without spreading at a flat-band point, which occurs at $\G=1$ and $\delta=\hbar\w/2$. Each bond is then an Ising coupling along the precessing axis, the A-site field vanishes in the rotating frame, and the Jordan--Wigner Majorana modes decouple into three-site clusters. For a uniform B-site field, the Bloch Hamiltonian becomes covariant, $\hh^\pm_k=e^{-ik\hat{\mathcal{K}}}\,\hh^\pm_0\,e^{ik\hat{\mathcal{K}}}$, where $\hat{\mathcal{K}}$ rotates the Nambu doublet on site A (End Matter~\ref{em:CLS}). The covariance carries over to $\hat U$, $\hat H_F$, and $\hat V$, so that all four bands are flat [Fig.~\ref{fig:4B_results_micro}(a)] and every term in Eq.~\eqref{eq:Aevolved} is $k$-independent in a suitable gauge. Equation~\eqref{eq:VarSplit} thus yields, for every band and at all times,
\begin{align}
    G^{kk}_n(t,t_0)=\mathrm{Var}_n[x](t),
    \label{eq:CLS-metric}
\end{align}
where the Wannier functions are the CLS of the flat bands~\cite{rhim2019classification, rhim2021singular, maimaiti2017compact, leykam2018artificial, rhim2020quantum, park2024quasilocalization}, i.e., Bogoliubov excitations supported on three consecutive sites [Eq.~\eqref{eq:CLS-op}]. Figure~\ref{fig:4B_results_micro}(b) shows both sides of Eq.~\eqref{eq:CLS-metric}, computed independently from the projector and from the real-space CLS profile. Here, $|\nu_0|=1$ and $\nu_\pi=0$, so the topological lower bounds at the two symmetric times coincide and do not require a metric change. Indeed, the chiral subspace is time independent, with $G_{\rm ch}=1$ throughout the drive cycle, and $G^{kk}_1+G^{kk}_2=G_{\rm ch}$ (End Matter~\ref{em:CLS}). The total CLS spread of bands 1 and 2 thus equals $G_{\rm ch}$, the lower bound for any Wannier basis of the chiral subspace, so these CLS are maximally localized Wannier functions (MLWFs) of the chiral subspace, and the micromotion only redistributes the fixed total spread between them.

Figure~\ref{fig:4B_results_micro}(c,d) contrasts the periodic breathing of a flat-band CLS at $\hbar\omega=2$ with the spreading of an MLWF in a dispersive Floquet band at $\hbar\omega=2.2$.

\smallskip
\textit{Experimental realization.---} The drive of Eq.~\eqref{eq:spinHam0} requires time-periodic, bond-resolved $\rm XX$/$\rm YY$ couplings with tunable anisotropy and $\s^x\s^y$ cross terms, and a staggered $\s^z$ field. Superconducting qubit arrays with tunable couplers provide this toolbox and have emulated flat-band lattices hosting CLS with site-resolved readout~\cite{rosen2025flatband}. In photonic lattices, CLS have been observed in Floquet flat bands~\cite{song2025observation} and with an oscillating internal weight distribution~\cite{hanafi2022localized}. The demanding ingredient is the anisotropic part, which generates Nambu pairing and requires parametric, excitation-nonconserving drives.

Equation~\eqref{eq:CLS-metric} furnishes a real-space probe of the band metric through the BdG weight $\rho^n_m(t)=|u^n_m(t)|^2+|v^n_m(t)|^2$, with particle and hole amplitudes $u^n_m$ and $v^n_m$, of the evolved CLS of band $n$ on site $m$ [Eq.~\eqref{eq:CLS-op}]. The Nambu doubling is an artifact of the Jordan--Wigner mapping rather than physical superconductivity, so the anomalous correlators correspond to spin correlators requiring no external phase reference. On the three-site CLS support, the Jordan--Wigner strings involve at most one intermediate $\s^z$, allowing local tomography to determine the normal and anomalous correlation matrix. Its diagonalization yields the BdG amplitudes $u^n_m(t)$ and $v^n_m(t)$, hence $\rho^n_m(t)$. Tracking this weight over one drive period directly reveals the flat-band breathing, $\mathrm{Var}_n[x](t)=G_n^{kk}(t)$ (SM).

\smallskip
\textit{Conclusion and outlook.---} Floquet driving not only shapes quasienergies and leads to stroboscopic Floquet topology but also impacts the quantum geometry of the Floquet states throughout the drive cycle. We have characterized the time evolution of the Floquet-state geometry within a unifying quantum geometric framework, revealing bounds on the size and rate of change of the integrated quantum metric beyond stroboscopic times. We have exemplified the theory for an exactly solvable chiral chain, where nontrivial $0$- and $\pi$-gap topology can enforce a finite change of the Floquet-state geometry during the drive. Because the integrated metric is the periodic lower bound on the Wannier spread, such changes can appear in real space as breathing, which becomes exact at a flat-band point, where the micromotion redistributes the site densities of the compact localized states, whose variances equal the band-resolved integrated quantum metrics.  

Our quantum geometric framework applies beyond the scope of the driven topological chain and enables straightforward generalizations beyond the quantum metric. The presented micromotion bounds hold for any isolated Floquet-band subspace in any dimension and make higher-dimensional flat bands natural settings for exploring geometric breathing, where the Wannier-spread relation generalizes to the position-covariance tensor~\cite{marzari1997}. We anticipate that the interplay between micromotion-operator geometry~\cite{he2026floquet} and Floquet-state geometry may yield new dynamical constraints in two-dimensional anomalous phases, where Floquet bands can be topologically trivial with topology residing in the micromotion~\cite{rudner2013anomalous}. The projector-based approach to Floquet states gives access to higher cumulants of the polarization distribution~\cite{Avdoshkin2023, Avdoshkin2025, Mitscherling2025} and to the temporal components of quantum geometry~\cite{dabiri2026time}. Since the ground-state metric constrains dissipation in slowly driven open systems~\cite{esin2025quantum}, extending the formulation to density matrices evolving under a master equation could reveal whether Floquet-state geometry also constrains heating, a major obstacle to stable Floquet engineering.

{\it Acknowledgment.---}  We thank Dan S. Borgnia, Marin Bukov, and Iliya Esin for valuable discussions. This work was supported by the Deutsche Forschungsgemeinschaft (DFG, German Research Foundation) through SFB 1170, the Würzburg-Dresden Cluster of Excellence ctd.qmat (EXC 2147, project-id 390858490), and the Emmy Noether Program (project-id 568440758).

\bibliography{refs_updated}

\section*{End matter}

\subsection{Branch cut, effective Hamiltonian, and Bloch convention}
\label{em:branchcut}

We define $\hat H_F=\frac{i\hbar}{T}\ln\hat U(\bk,t_0+T,t_0)$ with the branch cut of the logarithm in an open quasienergy gap. This makes $\hat H_F$ and the corresponding micromotion operator $\hat V$ smooth in $\bk$, with $\hat V(\bk,t_0+T,t_0)=\hat V(\bk,t_0,t_0)=\mathbbm{1}$. The micromotion operator is $T$-periodic and, for the square-wave drive of the main text, continuous and piecewise differentiable in time. Throughout the Letter, we place the cut in the $\pi$-gap, corresponding to the quasienergy zone $(-\hbar\omega/2,\hbar\omega/2]$. Moving the cut to another open gap shifts the quasienergies of the bands in between by $\hbar\omega$, so that $\hat V\to\hat V e^{i\omega(t-t_0)\hat P_C}$, where $\hat P_C$ projects onto these bands. This factor commutes with all Floquet-band projectors, so the Floquet-state geometry is independent of the cut, while $M^{aa}_S$, $G^{aa}_V$, and $K^{aa}_S$ are not. The micromotion bounds hold for any cut with an open gap (SM).

We Fourier transform with the actual site positions rather than the unit-cell index, $\hat H(\bk)=e^{-i\bk\cdot\hat{\mathbf X}}\hat H e^{i\bk\cdot\hat{\mathbf X}}$, with $\hat{\mathbf X}$ the site position operator. Bloch operators are then periodic up to conjugation by $e^{-i\mathbf b\cdot\hat{\mathbf X}}$ for reciprocal lattice vectors $\mathbf b$ ($b=\pi$ for our chain). Integrated metrics depend on this convention, and Eq.~\eqref{eq:VarSplit} requires it.

\subsection{Decomposition of the Floquet quantum metric}
\label{em:Gdecomposition}

For an isolated subspace $S$, we abbreviate $\hat P\equiv\hat P_S(\bk,t_0)$, $\hat{\bar P}\equiv\mathbbm{1}-\hat P$, and $\hat\xi\equiv\hat\xi^a_V(\bk,t,t_0)$. Differentiating Eq.~\eqref{eqn:timeFloquetProjector} yields $\hat V^\dagger(\partial_a\cP_S)\hat V=\partial_a\hat P-i[\hat\xi,\hat P]$. Inserting this into $g^{aa}_S=\frac12\Tr\big[(\partial_a\cP_S)^2\big]$ gives Eq.~\eqref{eq:g_decomposition}, with micromotion weight $\Tr\big[\hat P\hat\xi\hat{\bar P}\hat\xi\big]$ and interference term $\Tr\big[\hat{\mathcal L}^a_S\hat\xi\big]$, where $\hat{\mathcal L}^a_S=-i[\hat P,\partial_a\hat P]$. The momentum-off-diagonal components follow analogously (SM).

\subsection{Bound on the temporal variation of \texorpdfstring{$\sqrt{G^{aa}_S}$}{sqrt(GaaS)}}
\label{em:Mbound}

Integrating Eq.~\eqref{eq:g_decomposition} over the BZ gives $G^{aa}_S(t)=G^{aa}_S(t_0)+M(t)+I(t)$, where $M(t)\equiv M^{aa}_S(t,t_0)$ is the integral of the micromotion weight and $I(t)$ that of the interference term. Since $\hat P(\partial_a\hat P)\hat P=0$, the latter equals $2\,\mathrm{Im}\,\Tr[\hat P(\partial_a\hat P)\hat{\bar P}\hat\xi]$ before integration. The Hilbert-Schmidt Cauchy--Schwarz inequality, followed by a second Cauchy--Schwarz inequality over the BZ, gives
\begin{align}
    |I(t)| &\leq 2\sqrt{G^{aa}_S(t_0)M(t)}, \nonumber\\
    \left|\sqrt{G^{aa}_S(t_0)}-\sqrt{M(t)}\right|
    &\leq \sqrt{G^{aa}_S(t)} \leq \sqrt{G^{aa}_S(t_0)}+\sqrt{M(t)}.
    \label{eq:envelope}
\end{align}
The second line is equivalent to the first and implies the first bound in Eq.~\eqref{eq:micromotion-bound}. Unlike the latter, it still bounds $G^{aa}_S(t)$ from below when $M(t)>G^{aa}_S(t_0)$. The second bound in Eq.~\eqref{eq:micromotion-bound} follows from $\Tr[\hat\xi^2]\geq 2\Tr[\hat P\hat\xi\hat{\bar P}\hat\xi]$. Since the derivation uses only $\cP_S=\hat V\hat P\hat V^\dagger$, it holds equally for $\hat V\hat{\mathcal M}$ with any unitary $\hat{\mathcal M}$ commuting with $\hat P$, in particular for different branch cuts (SM).

\subsection{Derivation of the geometric speed limit}
\label{em:speedbound}

Inserting Eq.~\eqref{eqn:micromotionOperatorDefinition} into the Schr\"odinger equation gives $i\hbar\,\partial_t\cP_S=[\hat h,\cP_S]$, with $\hat h=i\hbar(\partial_t\hat V)\hat V^\dagger$, since $\cP_S$ commutes with $\hat V\hat H_F\hat V^\dagger$. Taking the momentum derivative and using the cyclicity of the trace yields
\begin{align}
    \dv{t}G^{aa}_S(t)
    =\frac{V_{\rm uc}}{(2\pi)^d\hbar}\int_{\rm BZ}\dd[d]{\bk}\,
    \Tr[\hat{\mathcal L}^a_S(t)\,\partial_a\hat h],
\end{align}
where $\hat{\mathcal L}^a_S(t)=-i[\cP_S,\partial_a\cP_S]$. The integrand has the same structure as the interference term above. Applying the same two Cauchy--Schwarz inequalities and using $\hbar\,\partial_t\hat\xi^a_V=\hat V^\dagger(\partial_a\hat h)\hat V$ gives $|\dv*{G^{aa}_S}{t}|\leq2\sqrt{G^{aa}_S K^{aa}_S}$, with $K^{aa}_S$ defined in the main text. Where $G^{aa}_S>0$, dividing by $2\sqrt{G^{aa}_S}$ yields Eq.~\eqref{eq:speed-limit}. Its integrated form also holds across zeros of the metric and the switching times of piecewise-smooth drives. The full derivation is given in the SM.

\subsection{Driven spin chain and chiral structure}
\label{em:chiralspinchain}

A Jordan--Wigner transformation and rotation to the frame of the precessing anisotropy axis yield Eq.~\eqref{eq:bloch_ham} in the Nambu basis $\Psi_k^\dagger=(A_k^\dagger,B_k^\dagger,A_{-k},B_{-k})$. The two Bloch Hamiltonians anticommute with $\Gamma=-\tau_y$, where $\tau_y$ acts in Nambu space. With $t_0=0$ and $\tau=T/2$, the half-period evolution operators satisfy $U_2=\Gamma U_1^\dagger\Gamma$ and $U=U_2U_1$. For a general time frame, a $\Gamma$-conjugate splitting requires $[U_t^\dagger\Gamma U_t\Gamma,U]=0$, with $U_t=U(k,t,t_0)$, which holds at the symmetric times but not generically elsewhere. The BdG particle-hole symmetry $\mathcal C=\tau_x\mathsf K$, together with $\Gamma$, places the model in class BDI. For a time-independent B-site field, $U_1=U_2$, so the two symmetric-time frames coincide and $\nu_\pi=0$ in the rotating frame. The transformation and rotating-frame conventions are detailed in the SM.

\subsection{Quantum-metric bound in chiral Floquet systems}
\label{em:chiralTopoBound}

In the chiral basis, the flattened effective Hamiltonian of either symmetric-time frame is $\mathbbm{1}-2\hat P_{\rm ch}$, with $\hat P_{\rm ch}=\frac12\big(\begin{smallmatrix}\mathbbm{1}&-\tilde q_k\\-\tilde q_k^\dagger&\mathbbm{1}\end{smallmatrix}\big)$ and $\tilde q_k$ an $\frac N2\times\frac N2$ unitary. With the Hermitian chiral connection $\Lambda_k=-i\tilde q_k^\dagger\partial_k\tilde q_k$, the winding number is $\nu=\frac{1}{2\pi}\int_{\rm BZ}\dd k\,\Tr\Lambda_k$, equal to $\nu_0+\nu_\pi$ at $t_0$ and $\nu_0-\nu_\pi$ at $t_0+\tau$~\cite{asboth2014chiral}, and $g_{\rm ch}(k)=\frac14\Tr(\Lambda_k^2)$. Applying the Cauchy--Schwarz inequality over the BZ, together with $(\Tr\Lambda_k)^2\leq\frac N2\Tr(\Lambda_k^2)$ and $V_{\rm BZ}V_{\rm uc}=2\pi$, yields $G_{\rm ch}\geq V_{\rm uc}^2\nu^2/(2N)$, i.e., Eq.~\eqref{eq:two-bounds}. Equality requires $\Lambda_k=\frac{2V_{\rm uc}\nu}{N}\mathbbm{1}_{N/2}$, a chiral connection that is both $k$-independent and proportional to the identity. Since $\Gamma$ acts within each site in our model, $\nu$ does not depend on the Bloch convention. The bound requires only chiral symmetry and holds for the one-dimensional chiral classes AIII, BDI, and CII~\cite{royharper2017periodic,yao2017}.

\subsection{Spread of a time-evolved Wannier function}
\label{em:wannierGt}

For a single isolated Floquet band $n$ in one dimension, we construct $\ket{W_n(t_0)}$ from the Bloch states $e^{ik\hat X}\ket{\varphi_n(k,t_0)}$ in a smooth, periodic gauge. Since the drive preserves lattice translations, the evolved state $\ket{W_n(t)}=\hat U(t,t_0)\ket{W_n(t_0)}$ is again a Wannier function, whose cell-periodic Bloch states are $\ket{u_n(k,t)}=e^{-i\varepsilon_n(k)(t-t_0)/\hbar}\ket{\chi_n(k,t,t_0)}$. We denote their Berry connection by $\mathcal A_n(k,t)=i\braket{u_n}{\partial_k u_n}$ and the BZ average by $\langle f\rangle_{\rm BZ}=\frac{V_{\rm uc}}{2\pi}\int_{\rm BZ}\dd k\,f(k)$.

The first two position moments are $\langle\hat X\rangle=\langle\mathcal A_n\rangle_{\rm BZ}$ and $\langle\hat X^2\rangle=\langle\braket{\partial_k u_n}{\partial_k u_n}\rangle_{\rm BZ}$. Using $g_n^{kk}=\braket{\partial_k u_n}{\partial_k u_n}-\mathcal A_n^2$, we obtain Eq.~\eqref{eq:VarSplit} with $\mathrm{Var}_k[f]=\langle f^2\rangle_{\rm BZ}-\langle f\rangle_{\rm BZ}^2\geq0$, and inserting $\ket{u_n(k,t)}$ into $\mathcal A_n$ gives Eq.~\eqref{eq:Aevolved}. Since $\hat V$ and $G_n^{kk}$ are $T$-periodic, the spread at fixed drive phase is quadratic in the cycle number $m$, with $m^2$ coefficient $T^2\mathrm{Var}_k[\partial_k\varepsilon_n]/\hbar^2$. For a quasienergy band periodic in $k$, $\langle\partial_k\varepsilon_n\rangle_{\rm BZ}=0$, so this coefficient equals $T^2\langle(\partial_k\varepsilon_n)^2\rangle_{\rm BZ}/\hbar^2$ and vanishes only for flat bands. The detailed cycle dependence and the gauge-dependent contribution are given in the SM.

If the band Wannier functions are compactly supported, as at a CLS point, the variance on an interval of length $L$ is at most $L^2/4$ (Popoviciu's inequality). Equation~\eqref{eq:width-bound} then requires the largest support length to obey $L\geq2V_{\rm uc}|\nu_s|/N$ site spacings. At the flat-band point, the actual value $G_{\rm ch}=1$ gives the stronger requirement $L\geq\sqrt2$ from Eq.~\eqref{eq:Var-chain}, i.e., at least two site spacings, which the three-site CLS saturates. For $|\nu_s|=1$ in our four-band chain, the topological bound alone requires only $L\geq1$.

\subsection{Flat bands, compact localized states, and their spatial variance}
\label{em:CLS}

At $\G=1$ and $\delta=\hbar\w/2$, the A-site Nambu doublet has no on-site terms, and Eq.~\eqref{eq:bloch_ham} becomes covariant, $\hh^\pm_k=e^{-ik\hat{\mathcal K}}\hh^\pm_0e^{ik\hat{\mathcal K}}$ with $\hat{\mathcal K}=i(\hat e_3\hat e_1^\top-\hat e_1\hat e_3^\top)$. This covariance carries over to $\hat U$, $\hat H_F$, and $\hat V$, so that all bands are flat. At $\G=\delta=\hbar\w/2=1$, the vectors $w_1(k)=(\sin k,0,-\cos k,0)^\top$ and $w_2=(0,0,0,1)^\top$ satisfy $\hh^\pm_kw_1=w_2$ and $\hh^\pm_kw_2=w_1+(\frac{\hbar\w}{2}\pm\delta)w_2$. Their span is invariant and contains the two negative-quasienergy bands, whose states evolve as $a_n(t)w_1(k)+b_n(t)w_2$ with $k$-independent coefficients and $|a_n|^2+|b_n|^2=1$.

Hence $\hat P_{\rm ch}(k,t)=w_1w_1^\top+w_2w_2^\top$ is time independent, so $G_{\rm ch}=1$. Its chiral connection has eigenvalues $\{0,2\}$, giving $\nu_0=1$ and $\nu_\pi=0$. Since $w_1$ is real and normalized and $w_2\perp\partial_kw_1$, we obtain $g^{kk}_n=|a_n(t)|^2$ and $G^{kk}_1+G^{kk}_2=G_{\rm ch}$. The complementary invariant plane spanned by $\Gamma w_1$ and $\Gamma w_2$ yields the same metric-variance relation for the remaining two bands.

Following Ref.~\cite{rhim2019classification}, superposing the negative-quasienergy flat-band Bloch states with envelope $e^{ik}$ yields a CLS on three consecutive sites, with Bogoliubov operator
\begin{align}
    -\tfrac{a_n}{2}\big(if^\dagger_{2\ell-2}+f_{2\ell-2}\big)
    +b_n f_{2\ell-1}
    +\tfrac{a_n}{2}\big(if^\dagger_{2\ell}-f_{2\ell}\big),
    \label{eq:CLS-op}
\end{align}
where $a_n=a_n(t)$ and $b_n=b_n(t)$. The BdG weight $\rho^n_m=|u^n_m|^2+|v^n_m|^2$ is $|a_n|^2/2$ on sites $2\ell-2$ and $2\ell$ and $|b_n|^2$ on site $2\ell-1$. Thus the center stays at $2\ell-1$ and $\mathrm{Var}_n[x](t)=|a_n(t)|^2=G_n^{kk}(t,t_0)$. The CLS remains on its three-site support while its internal weight distribution breathes, and that breathing is exactly the time-resolved metric. The particle-hole partners provide CLS of the remaining two bands. The explicit flat-band eigenvectors and coefficients are given in the SM.

\end{document}


\allowdisplaybreaks

\title{Supplemental Material for ``Micromotion and topological constraints on Floquet-state quantum geometry''}

\author{Arpit Raj}
\affiliation{Institute for Theoretical Physics and Astrophysics, University of W\"urzburg, D-97074 W\"urzburg, Germany}
\affiliation{W\"urzburg-Dresden Cluster of Excellence ctd.qmat, D-97074 W\"urzburg, Germany}

\author{Johannes Mitscherling}
\affiliation{Institute for Theoretical Physics and Astrophysics, University of W\"urzburg, D-97074 W\"urzburg, Germany}
\affiliation{W\"urzburg-Dresden Cluster of Excellence ctd.qmat, D-97074 W\"urzburg, Germany}
\affiliation{Max Planck Institute for the Physics of Complex Systems, N\"othnitzer Str. 38, 01187 Dresden, Germany}

\author{Bj\"orn Trauzettel}
\affiliation{Institute for Theoretical Physics and Astrophysics, University of W\"urzburg, D-97074 W\"urzburg, Germany}
\affiliation{W\"urzburg-Dresden Cluster of Excellence ctd.qmat, D-97074 W\"urzburg, Germany}

\maketitle


\section{Branch cut, effective Hamiltonian, and Bloch convention}
\label{sm:branchcut}

We supplement End Matter~\ref{mt:em:branchcut} with the explicit branch-cut construction, the transformation of the micromotion connection under a change of cut, and the action of the position operator in the site-position Bloch convention.

The effective Floquet Hamiltonian with the cut at $e^{-i\epsilon}$, corresponding to the quasienergy $\varepsilon=\epsilon\hbar/T$, is $\hat H_{F}^{(\epsilon)}(\bk)=\frac{i\hbar}{T}\ln^{\epsilon} \hat U_F(\bk)$, with $\hat U_F(\bk)=\hat U(\bk,t_0+T,t_0)$. It uses the argument
\begin{align}
    \arg^\epsilon(z)
    =\mathrm{mod}\big(\arg z+\epsilon,\,2\pi\big)-\epsilon,
\end{align}
with output in $[-\epsilon,2\pi-\epsilon)$. Writing
$\hat U_F(\bk) = \sum_n e^{-i\varepsilon_n(\bk)T/\hbar} \ket{\varphi_n(\bk)}\,\bra{\varphi_n(\bk)}$, one obtains
\begin{align}
    \hat H_{F}^{(\epsilon)}(\bk) = -\frac{\hbar}{T}
    \sum_n \arg^\epsilon\,\big( e^{-i\varepsilon_n(\bk)T/\hbar} \big)
    \ket{\varphi_n(\bk)}\,\bra{\varphi_n(\bk)} .
\end{align}
The $\pi$-gap cut used throughout is $\epsilon=\pi$, and we drop the label $\epsilon$.

Moving the cut to $\epsilon'>\epsilon$ in another open gap gives $\hat H_F^{(\epsilon')}=\hat H_F^{(\epsilon)}+\hbar\w\,\hat P_C$ and hence $\hat V_{\epsilon'}=\hat V_\epsilon\,e^{i\omega(t-t_0)\hat P_C}$, where $\hat P_C$ projects onto the bands between the two cuts; for example, $\hat P_C=\hat P_{\rm ch}$ when moving the cut from the $\pi$- to the $0$-gap. The micromotion connection transforms as
\begin{align}
    \hat\xi^a_{V_{\epsilon'}}
    = e^{-i\omega(t-t_0)\hat P_C}\,\hat\xi^a_{V_\epsilon}\,e^{i\omega(t-t_0)\hat P_C}
    + i\,e^{-i\omega(t-t_0)\hat P_C}\,\partial_a e^{i\omega(t-t_0)\hat P_C} ,
\end{align}
where the second term is nonzero when $\hat P_C$ depends on $\bk$, except at the stroboscopic times $t=t_0+mT$. Hence $M^{aa}_S$, $G^{aa}_V$, and $K^{aa}_S$ depend on the cut, while Eqs.~\eqref{mt:eq:micromotion-bound}, \eqref{mt:eq:speed-limit}, and \eqref{mt:eq:shared} hold for any cut with an open gap (Secs.~\ref{sm:Mbound} and \ref{sm:speedbound}).

In the site-position Bloch convention of End Matter~\ref{mt:em:branchcut}, a state $\ket{\psi}=\frac{V_{\rm uc}}{(2\pi)^d}\int_{\rm BZ}\dd[d]{\bk}\,e^{i\bk\cdot\hat{\mathbf X}}\ket{u(\bk)}$ has components obeying $\ket{u(\bk+\mathbf b)}=e^{-i\mathbf b\cdot\hat{\mathbf X}}\ket{u(\bk)}$, so that the integrand is periodic over the BZ. Using $\hat X_a e^{i\bk\cdot\hat{\mathbf X}}=-i\partial_a e^{i\bk\cdot\hat{\mathbf X}}$ and integrating by parts without boundary terms gives
\begin{align}
    \hat X_a\ket{\psi}=\frac{V_{\rm uc}}{(2\pi)^d}\int_{\rm BZ}\dd[d]{\bk}\,e^{i\bk\cdot\hat{\mathbf X}}\,i\partial_a\ket{u(\bk)} ,
\end{align}
i.e., $\hat X_a$ acts as $i\partial_{k_a}$ on the Bloch components. For the micromotion operator, which acts on them as $\hat V(\bk,t,t_0)$, this yields
\begin{align}
    \hat V^\dagger\hat X_a\hat V
    = \hat X_a + i\,\hat V^\dagger\partial_a\hat V
    = \hat X_a+\hat\xi^a_V ,
    \label{sm:eq:Xshift}
\end{align}
which gives the micromotion-induced position displacement quoted in the main text. With the unit-cell index in place of the site positions, $i\partial_{k_a}$ would instead represent the unit-cell position, and the Bloch projectors would differ by the $\bk$-dependent conjugation $e^{i\bk\cdot\hat{\boldsymbol\tau}}$, with $\hat{\boldsymbol\tau}$ the intracell positions, which changes the metric.

\section{Decomposition of the Floquet quantum metric}
\label{sm:Gdecomposition}

We derive the momentum components of the quantum metric of the time-evolved Floquet projector $\cP_S(\bk,t,t_0)$ for a projector $\hat P_S(\bk,t_0)$ onto any isolated set $S$ of Floquet bands, which includes a single band $S=\{n\}$. We abbreviate $\hat P\equiv\hat P_S(\bk,t_0)$, $\hat{\bar P}\equiv\mathbbm{1}-\hat P$, and $\hat V\equiv\hat V(\bk,t,t_0)$. Differentiating $\cP_S=\hat V\hat P\hat V^\dagger$ [Eq.~\eqref{mt:eqn:timeFloquetProjector}] and using the micromotion connection $\hat\xi^a_V=i\,\hat V^\dagger\partial_a\hat V$, together with $(\partial_a\hat V^\dagger)\hat V=-\hat V^\dagger\partial_a\hat V$, gives
\begin{align}
    \hat V^\dagger\big(\partial_a\cP_S\big)\hat V=\partial_a\hat P-i\big[\hat\xi^a_V,\hat P\big] ,
    \label{sm:eq:dPconj}
\end{align}
which directly separates the stroboscopic derivative from the micromotion contribution. Inserting this identity into $g^{ab}_S=\frac12\Tr[(\partial_a\cP_S)(\partial_b\cP_S)]$ yields
\begin{align}
    g^{ab}_S(\bk,t,t_0) = \frac{1}{2}g^{ab}_S(\bk,t_0)
    +\frac{1}{2}\Tr\big[\hat P\,\hat\xi^a_V\,\hat{\bar P}\,\hat\xi^b_V\big]
    +\frac{1}{2}\Tr\big[\hat{\mathcal L}^a_S\,\hat\xi^b_V\big] + (a\leftrightarrow b) ,
    \label{sm:eq:gab}
\end{align}
where $(a\leftrightarrow b)$ denotes the symmetrization of all terms in the indices $a$ and $b$, and $\hat{\mathcal L}^a_S=-i\big[\hat P,\partial_a\hat P\big]$ is the subspace adiabatic connection operator~\cite{Mitscherling2025Orbital} of the Floquet projector. The square of the first term in Eq.~\eqref{sm:eq:dPconj} gives the stroboscopic metric, the product of the two commutators gives the micromotion weight, $-\frac12\Tr\big[[\hat\xi^a_V,\hat P][\hat\xi^b_V,\hat P]\big]=\frac12\Tr[\hat P\hat\xi^a_V\hat{\bar P}\hat\xi^b_V]+(a\leftrightarrow b)$, and the cross terms give the interference term. For $a=b$, Eq.~\eqref{sm:eq:gab} reduces to Eq.~\eqref{mt:eq:g_decomposition}. Since the derivation uses only $\hat P^2=\hat P$, it applies equally to individual bands, with $\hat{\mathcal L}^a_n=-i\big[\hat P_n,\partial_a\hat P_n\big]$ the band-resolved adiabatic connection operator, and to isolated multiband subspaces. By Eq.~\eqref{sm:eq:Xshift}, the commutator in Eq.~\eqref{sm:eq:dPconj} is the change of $\partial_a\hat P=-i[\hat X_a,\hat P]$ under the micromotion-induced displacement $\hat\xi^a_V$ of the position operator.

\section{Bound on the temporal variation of \texorpdfstring{$\sqrt{G^{aa}_S}$}{sqrt(GaaS)}}
\label{sm:Mbound}

We derive a bound on the variation of the integrated Floquet quantum metric from the diagonal components of Eq.~\eqref{sm:eq:gab}. Throughout, $a$ labels a momentum direction $k_a$. For compactness, we abbreviate $\hat P\equiv\hat P_S(\vb{k},t_0)$, $\hat{\bar P}\equiv\mathbbm{1}-\hat P$, $\hat\xi\equiv\hat\xi^a_V(\vb{k},t,t_0)$, and $\hat{\mathcal{L}} \equiv \hat {\mathcal{L}}^a_S(\vb{k},t_0) = -i\comm{\hat P}{\partial_a\hat P}$. Integrating Eq.~\eqref{sm:eq:gab} at $a=b$ over the Brillouin zone yields
\begin{align}
    G^{aa}_S(t) &= G^{aa}_S(t_0) + M(t) + I(t),
    \label{sm:eq:Gtsplit}
\end{align}
where the micromotion contribution $M(t)\equiv M^{aa}_S(t,t_0)$ of the main text and the interference contribution are
\begin{align}
M(t) &= \frac{V_\mathrm{uc}}{(2\pi)^d} \int_{\rm BZ}\dd[d]{k}\,
\Tr[ \hat P\hat\xi\hat{\bar P}\hat\xi ],
\label{sm:eq:Mdef}\\
I(t) &= \frac{V_\mathrm{uc}}{(2\pi)^d} \int_{\rm BZ}\dd[d]{k}\, \Tr[ \hat{\mathcal{L}}\hat\xi ].
\label{sm:eq:Idef}
\end{align}
The micromotion contribution is nonnegative, since $\Tr[\hat P\hat\xi\hat{\bar P}\hat\xi]=\|\hat P\hat\xi\hat{\bar P}\|_{\rm HS}^2\geq0$, with the Hilbert--Schmidt norm $\|\hat{\mathcal O}\|_{\rm HS}^2=\Tr[\hat{\mathcal O}^\dag\hat{\mathcal O}]$. Using $\hat P(\partial_a\hat P)\hat P=\hat{\bar P}(\partial_a\hat P)\hat{\bar P}=0$ and the Hermiticity of $\partial_a\hat P$ and $\hat\xi$, the interference integrand becomes $\Tr[\hat{\mathcal L}\hat\xi]=-2\,\mathrm{Im}\,\Tr[(\hat{\bar P}\,\partial_a\hat P\,\hat P)(\hat P\,\hat\xi\,\hat{\bar P})]$. The Cauchy--Schwarz inequality for the Hilbert--Schmidt inner product, together with $\Tr[\hat P(\partial_a\hat P)\hat{\bar P}(\partial_a\hat P)]=\frac12\Tr[(\partial_a\hat P)^2]=g^{aa}_S(\vb{k},t_0)$, gives
\begin{align}
\big|\Tr[\hat{\mathcal{L}}\,\hat\xi]\big| \leq 2\sqrt{g^{aa}_S(\vb{k},t_0)}\,\sqrt{\Tr[\hat P\,\hat\xi\,\hat{\bar P}\,\hat\xi]} ,
\end{align}
and a second Cauchy--Schwarz inequality over the Brillouin zone gives
\begin{align}
|I(t)| \leq 2\sqrt{G^{aa}_S(t_0)}\,\sqrt{M(t)} .
\end{align}
Inserting this into Eq.~\eqref{sm:eq:Gtsplit} bounds $G^{aa}_S(t)$ between $\big(\sqrt{G^{aa}_S(t_0)}\mp\sqrt{M(t)}\big)^2$, so that $\sqrt{G^{aa}_S(t)}$ obeys the two-sided envelope
\begin{align}
\left|\sqrt{G^{aa}_S(t_0)}-\sqrt{M(t)}\right| \leq \sqrt{G^{aa}_S(t)} \leq \sqrt{G^{aa}_S(t_0)}+\sqrt{M(t)} ,
\label{sm:eq:envelope}
\end{align}
which yields the bound
\begin{align}
\Big|\sqrt{G^{aa}_S(t)}-\sqrt{G^{aa}_S(t_0)}\,\Big| \leq \sqrt{M(t)}.
\label{sm:eq:Mbound}
\end{align}
The second, subspace-independent inequality in Eq.~\eqref{mt:eq:micromotion-bound} follows from $\Tr[\hat P\hat\xi\hat{\bar P}\hat\xi] \leq \frac{1}{2} \Tr [\hat\xi \hat\xi ] = \frac{1}{2} \Tr [\partial_a \hat V^\dag \partial_a \hat V ]$, since $\Tr[\hat\xi\hat\xi]=2\Tr[\hat P\hat\xi\hat{\bar P}\hat\xi]+\|\hat P\hat\xi\hat P\|_{\rm HS}^2+\|\hat{\bar P}\hat\xi\hat{\bar P}\|_{\rm HS}^2$.

The derivation relies on $\cP_S(\vb{k},t,t_0)=\hat V\hat P\hat V^\dag$. It therefore holds equally for $\hat V\hat{\mathcal M}$ with any unitary $\hat{\mathcal M}$ commuting with $\hat P$, in particular for the micromotion operator of any branch cut whose gap is open (End Matter~\ref{mt:em:branchcut}), for which $\hat{\mathcal M}=e^{i\omega(t-t_0)\hat P_C}$ (Sec.~\ref{sm:branchcut}).

\section{Derivation of the geometric speed limit}
\label{sm:speedbound}

We abbreviate $\hat V\equiv\hat V(\vb{k},t,t_0)$. Inserting Eq.~\eqref{mt:eqn:micromotionOperatorDefinition} into $i\hbar\,\partial_t\hat U(\vb{k},t,t_0)=\hat H(\vb{k},t) \hat U(\vb{k},t,t_0)$ yields
\begin{align}
    \hat H(\vb{k},t) &= \hat h(\vb{k},t,t_0) + \hat V\hat H_F(\vb{k},t_0)\hat V^\dag,
    \label{sm:eq:Hsplit}
\end{align}
where
\begin{align}
    \hat h(\vb{k},t,t_0) &\equiv i\hbar\,\big(\partial_t\hat V\big)\,\hat V^\dag,
\end{align}
which is Hermitian since $\partial_t(\hat V\hat V^\dag)=0$. In this section, we abbreviate $\hat P(t)\equiv \cP_S(\vb{k},t,t_0)$ and $\hat P(t_0)\equiv\hat P_S(\vb{k},t_0)$ (and accordingly their complements). Since $\comm{\hat P(t_0)}{\hat H_F(\vb{k},t_0)}=0$ and $\hat P(t)=\hat V\hat P(t_0)\hat V^\dag$ by Eq.~\eqref{mt:eqn:timeFloquetProjector}, we have $\comm{\hat P(t)}{\hat V\hat H_F(\vb{k},t_0)\hat V^\dag}=0$. Thus, $i\hbar\,\partial_t\hat P(t) = \comm{\hat H}{\hat P(t)} = \comm{\hat h}{\hat P(t)}$.
Taking the momentum derivative gives
\begin{align}
    i\hbar\,\partial_t\partial_a\hat P(t) = \comm{\partial_a\hat h}{\hat P(t)} + \comm{\hat h}{\partial_a\hat P(t)}.
\end{align}
The time derivative of the integrated Floquet quantum metric then reads
\begin{align}
    \dv{t}G^{aa}_S(t)
    &= \frac{V_\mathrm{uc}}{(2\pi)^d} \int_{\rm BZ}\dd[d]{k}\, \Tr[(\partial_a\hat P(t))\,\partial_t\partial_a\hat P(t)] \nonumber\\
    &= -\frac{i}{\hbar}\frac{V_\mathrm{uc}}{(2\pi)^d} \int_{\rm BZ}\dd[d]{k}\, \Tr[(\partial_a\hat P(t))\comm{\partial_a\hat h}{\hat P(t)}] \nonumber\\
    &= \frac{1}{\hbar}\frac{V_\mathrm{uc}}{(2\pi)^d} \int_{\rm BZ}\dd[d]{k}\, \Tr[\hat{\mathcal{L}}(t)\,\partial_a\hat h].
    \label{sm:eq:dGdt}
\end{align}
In the second line, the contribution of $\comm{\hat h}{\partial_a\hat P(t)}$ vanishes, since $\Tr[(\partial_a\hat P(t))\comm{\hat h}{\partial_a\hat P(t)}]=0$ by the cyclicity of the trace. In the last line, we used $\Tr[(\partial_a\hat P(t))\comm{\partial_a\hat h}{\hat P(t)}]=\Tr[(\partial_a\hat h)\comm{\hat P(t)}{\partial_a\hat P(t)}]$.

The integrand in Eq.~\eqref{sm:eq:dGdt} has the same structure as the interference contribution $\Tr[\hat{\mathcal{L}}\hat\xi]$ of the previous section, with $\hat P\to\hat P(t)$ and $\hat\xi\to\partial_a\hat h$. The steps used there only required $\hat P^2=\hat P$ and the Hermiticity of $\hat\xi$. Hence,
\begin{align}
    \big|\Tr[\hat{\mathcal{L}}(t)\,\partial_a\hat h]\big|
    &\leq 2\sqrt{g^{aa}_S(\vb{k},t,t_0)}\,\sqrt{\Tr[\hat P(t)\,(\partial_a\hat h)\,\hat{\bar P}(t)\,(\partial_a\hat h)]}.
\end{align}
Applying the Cauchy--Schwarz inequality to the Brillouin-zone integral gives
\begin{align}
    \bigg|\dv{t}G^{aa}_S(t)\bigg| \leq 2\sqrt{G^{aa}_S(t)}\,\sqrt{K^{aa}_S(t)},
    \label{sm:eq:dGbound}
\end{align}
with
\begin{align}
    K^{aa}_S(t) &= \frac{1}{\hbar^2} \frac{V_\mathrm{uc}}{(2\pi)^d} \int_{\rm BZ}\dd[d]{k}\,
    \Tr[ \hat P(t)\,(\partial_a\hat h)\,\hat{\bar P}(t)\,(\partial_a\hat h) ].
\end{align}
Using $\frac{1}{2\sqrt{G^{aa}_S}}\dv{}{t}G^{aa}_S = \dv{}{t}\sqrt{G^{aa}_S}$ wherever $G^{aa}_S(t)\neq0$, we obtain the geometric speed limit
\begin{align}
    \bigg|\dv{t}\sqrt{G^{aa}_S(t)}\,\bigg| \leq \sqrt{K^{aa}_S(t)}.
    \label{sm:eq:speedlimit}
\end{align}
Furthermore, since $i\hbar\,\partial_t\hat V=\hat h\hat V$, it follows that $\hbar\,\partial_t\hat\xi^a_V=\hat V^\dag(\partial_a\hat h)\hat V$. Together with $\hat P(t)=\hat V\hat P(t_0)\hat V^\dag$, this gives $\Tr[\hat P(t)\,(\partial_a\hat h)\,\hat{\bar P}(t)\,(\partial_a\hat h)]=\hbar^2\,\Tr[\hat P(t_0)\,(\partial_t\hat\xi^a_V)\,\hat{\bar P}(t_0)\,(\partial_t\hat\xi^a_V)]$. Thus,
\begin{align}
    K^{aa}_S(t) &= \frac{V_\mathrm{uc}}{(2\pi)^d} \int_{\rm BZ}\dd[d]{k}\,
    \Tr[ \hat P(t_0)\,(\partial_t\hat\xi^a_V)\,\hat{\bar P}(t_0)\,(\partial_t\hat\xi^a_V)].
    \label{sm:eq:Kxi}
\end{align}
As in Sec.~\ref{sm:Mbound}, the derivation only uses $\cP_S=\hat V\hat P\hat V^\dag$ and therefore holds for the micromotion operator of any branch cut whose gap is open, with the corresponding $K^{aa}_S$.

\textit{Integrated form.---} Equation~\eqref{sm:eq:speedlimit} requires $G^{aa}_S\neq0$ and a differentiable metric. For a piecewise-smooth drive, $G^{aa}_S$ is continuous and piecewise differentiable, while $K^{aa}_S$ jumps at the switching times but remains bounded, so integrating Eq.~\eqref{sm:eq:speedlimit} between switching times gives
\begin{align}
    \Big|\sqrt{G^{aa}_S(t_2)}-\sqrt{G^{aa}_S(t_1)}\,\Big| \leq \int_{t_1}^{t_2}\dd{t}\,\sqrt{K^{aa}_S(t)} ,
    \qquad t_1\leq t_2 ,
    \label{sm:eq:speedlimit-int}
\end{align}
whenever $G^{aa}_S>0$ in $(t_1,t_2)$. If $G^{aa}_S$ vanishes in $[t_1,t_2]$, applying this bound between the last zero and $t_2$, or between $t_1$ and the first zero, bounds the larger of $\sqrt{G^{aa}_S(t_1)}$ and $\sqrt{G^{aa}_S(t_2)}$, and hence their difference.

\textit{Velocity bound.---} For any Hermitian $\hat{\mathcal O}$, $\Tr[\hat P\hat{\mathcal O}\hat{\bar P}\hat{\mathcal O}]=\|\hat{\bar P}\hat{\mathcal O}\hat P\|^2_{\rm HS}\leq\mathrm{rank}(\hat P)\,\|\hat{\mathcal O}\|^2_{\rm op}$. Applied to Eq.~\eqref{sm:eq:Kxi} with $\mathrm{rank}(\hat P)=N_S$ and the normalization $\frac{V_{\rm uc}}{(2\pi)^d}\int_{\rm BZ}\dd[d]{k}=1$, this gives
\begin{align}
    \sqrt{K^{aa}_S(t)}\leq\sqrt{N_S}\,v^a_{\max},
    \qquad
    v^a_{\max}=\max_{\vb{k},t}\big\|\partial_t\hat\xi^a_V\big\|_{\rm op}
    =\frac{1}{\hbar}\max_{\vb{k},t}\big\|\partial_a\hat h\big\|_{\rm op} ,
\end{align}
which reduces to the single-band bound of the main text for $N_S=1$.

\section{Driven spin chain and chiral structure}
\label{sm:chiralspinchain}

A Jordan--Wigner transformation, Fourier transform, and rotation to the frame of the precessing anisotropy axis yield Eq.~\eqref{mt:eq:bloch_ham} in the Nambu basis $\Psi_k^\dagger=(A_k^\dagger,B_k^\dagger,A_{-k},B_{-k})$. Here, we use $\s^z_m=1-2f^\dagger_mf_m$ and $\s^+_m=e^{i\pi\sum_{j<m}f^\dagger_jf_j}f_m$, where $\s^\pm_m=\tfrac12(\s^x_m\pm i\s^y_m)$, with $A_\ell\equiv f_{2\ell}$ and $B_\ell\equiv f_{2\ell-1}$, the site-resolved Fourier transform $A_k=N_{\rm uc}^{-1/2}\sum_\ell e^{-2ik\ell}A_\ell$ and $B_k=N_{\rm uc}^{-1/2}\sum_\ell e^{-ik(2\ell-1)}B_\ell$ for $N_{\rm uc}$ unit cells, and the rotation $\hat R(t)=e^{i\w t\hat N/2}$ with $\hat N=\sum_m f^\dagger_mf_m$. We now detail these steps.

\textit{Spin Hamiltonian.---} In terms of $\s^\pm_m$, Eq.~\eqref{mt:eq:spinHam0} reads
\begin{align}
    H(t)=\sum_m\Big[\tfrac12\big(\s^+_m\s^-_{m+1}+\s^-_m\s^+_{m+1}\big)
    +\tfrac{i\G}{2}\big(e^{i\w t}\s^+_m\s^+_{m+1}-e^{-i\w t}\s^-_m\s^-_{m+1}\big)
    -\frac{\delta_m(t)}{2}\,\s^z_m\Big] ,
    \label{sm:eq:Hspin-pm}
\end{align}
i.e., isotropic $\rm XY$ exchange and a uniaxial anisotropy $\G$ whose axis precesses in the $xy$ plane at $\w/2$.

\textit{Jordan--Wigner transformation.---} The string operators cancel in all nearest-neighbor products, $\s^+_m\s^-_{m+1}=f^\dagger_{m+1}f_m$ and $\s^+_m\s^+_{m+1}=f_{m+1}f_m$, so that Eq.~\eqref{sm:eq:Hspin-pm} maps onto
\begin{align}
    H(t)=\sum_m\Big[\tfrac12\big(f^\dagger_mf_{m+1}+f^\dagger_{m+1}f_m\big)
    +\tfrac{i\G}{2}\big(e^{i\w t}f_{m+1}f_m-e^{-i\w t}f^\dagger_mf^\dagger_{m+1}\big)
    +\delta_m(t)\,f^\dagger_mf_m\Big] ,
    \label{sm:eq:Hfermion-lab}
\end{align}
up to a $c$-number. The mapping is exact for an open chain; for a ring, it adds a boundary term that depends on the fermion parity but does not affect the bulk.

\textit{Rotating frame.---} The states in the frame of the precessing axis are $|\tilde\psi(t)\rangle=\hat R(t)\ket{\psi(t)}$ and evolve under $\tilde H(t)=\hat R H\hat R^\dagger+i\hbar(\partial_t\hat R)\hat R^\dagger=\hat RH\hat R^\dagger-\tfrac{\hbar\w}{2}\hat N$. Since $\hat Rf_m\hat R^\dagger=e^{-i\w t/2}f_m$, the pairing phases in Eq.~\eqref{sm:eq:Hfermion-lab} cancel, and
\begin{align}
    \tilde H(t)=\sum_m\Big[\tfrac12\big(f^\dagger_mf_{m+1}+f^\dagger_{m+1}f_m\big)
    +\tfrac{i\G}{2}\big(f_{m+1}f_m-f^\dagger_mf^\dagger_{m+1}\big)
    +\big(\delta_m(t)-\tfrac{\hbar\w}{2}\big)f^\dagger_mf_m\Big] .
    \label{sm:eq:Hfermion-rot}
\end{align}
The bond terms are static, the on-site terms are offset by $\hbar\w/2$, and only the longitudinal field on the B sites remains driven, $\delta_{2\ell-1}(t)=\mp\delta$ for $\cos\w t\gtrless0$. In spin language, $\hat R(t)$ rotates all spins about the $z$ axis by the angle $\w t/2$, which removes the phases $e^{\pm i\w t}$ in Eq.~\eqref{sm:eq:Hspin-pm} and replaces $\delta_m(t)$ by $\delta_m(t)-\hbar\w/2$. The anisotropy axis is then static along $(\mathbf e_x-\mathbf e_y)/\sqrt2$, and for $\G=1$ each bond is an Ising coupling along it.

\textit{Bloch Hamiltonian.---} Inserting $f_{2\ell}=N_{\rm uc}^{-1/2}\sum_ke^{2ik\ell}A_k$ and $f_{2\ell-1}=N_{\rm uc}^{-1/2}\sum_ke^{ik(2\ell-1)}B_k$ with $k\in[-\tfrac\pi2,\tfrac\pi2)$ into Eq.~\eqref{sm:eq:Hfermion-rot} gives
\begin{align}
    \tilde H&=\sum_k\Big[\cos k\,\big(A^\dagger_kB_k+B^\dagger_kA_k\big)
    +\G\sin k\,\big(A_{-k}B_k+B^\dagger_kA^\dagger_{-k}\big)
    +\big(\delta-\tfrac{\hbar\w}{2}\big)A^\dagger_kA_k
    +\big(\mp\delta-\tfrac{\hbar\w}{2}\big)B^\dagger_kB_k\Big] \nonumber\\
    &=\frac12\sum_k\Psi^\dagger_k\,\hh^\pm_k\,\Psi_k ,
\end{align}
up to a $c$-number, with $\hh^\pm_k$ of Eq.~\eqref{mt:eq:bloch_ham}. Here, each bond between sites $m$ and $m+1$ contributes the phase factors $e^{\pm ik}$ irrespective of the sublattice of $m$, which produces the $\cos k$ and $\sin k$ entries. In this convention (End Matter~\ref{mt:em:branchcut}), $\hh^\pm_{k+\pi}=\hat D\hh^\pm_k\hat D$ with $\hat D=\mathrm{diag}(1,-1,1,-1)$.

\textit{Symmetric times.---} With the chiral operator $\Gamma=-\tau_y$, where $\tau_i$ are Pauli matrices in Nambu space, $t_0=0$, and $\tau=T/2$, the half-period evolution operators are
\begin{align}
    U_1=e^{-\frac{i}{\hbar}\frac{T}{4}\hh^-_k}\,e^{-\frac{i}{\hbar}\frac{T}{4}\hh^+_k},
    \qquad
    U_2=e^{-\frac{i}{\hbar}\frac{T}{4}\hh^+_k}\,e^{-\frac{i}{\hbar}\frac{T}{4}\hh^-_k}=\Gamma U_1^\dagger\Gamma ,
\end{align}
so that $U=U_2U_1$.

\textit{Rotating versus laboratory frame.---} The rotating-frame transformation shifts all quasienergies by $\hbar\w/2$, interchanging the $0$- and $\pi$-gaps. Both $(\nu_0+\nu_\pi)^2$ and $(\nu_0-\nu_\pi)^2$ are invariant under that exchange, so Eq.~\eqref{mt:eq:two-bounds} is unaffected. Since $\hat R(t)$ is $k$-independent, the Floquet-state geometry, including $G_{\rm ch}(t)$, is identical in both frames, whereas the micromotion operators of the two frames are related by the $k$-independent rotation $\hat R(t)$ and a factor accounting for the relabeling of the quasienergy zone. The rotation drops out of $\hat\xi^a_V$, so $M^{aa}_S$, $G^{aa}_V$, and $K^{aa}_S$ change only through the branch-cut factor of End Matter~\ref{mt:em:branchcut}. Throughout the paper, we quote the topological invariants and evaluate $M^{aa}_S$, $G^{aa}_V$, and $K^{aa}_S$ in the rotating frame.

\textit{Uniform static field.---} For $\delta_m(t)=\delta$ on all sites, the chain has one-site translation symmetry, and Eq.~\eqref{sm:eq:Hfermion-lab} reduces to the two-band BdG Hamiltonian $h_{\rm lab}(k,t)=(\cos k+\delta)\tau_z+\G\sin k\,(\cos\w t\,\tau_x+\sin\w t\,\tau_y)$ in the Nambu basis $(f_k,f^\dagger_{-k})$, the harmonically driven model of Ref.~\cite{yang2019floquet} with $\delta_1=1$, $\delta_2=\delta$, and $\Omega=\G$, up to the normalization of the Nambu form.

\section{Spread of a time-evolved Wannier function}
\label{sm:wannierGt}

We use the notation of End Matter~\ref{mt:em:wannierGt} for a single isolated Floquet band $n$ in one dimension. With the quasienergy gap at the branch cut open, $\varepsilon_n(k)$ is smooth and periodic in $k$, and $\hat V(k,t,t_0)$ is smooth in $k$ and obeys the Bloch boundary condition of End Matter~\ref{mt:em:branchcut}.

\textit{Cycle dependence.---} With the group velocity $v_n(k)=\partial_k\varepsilon_n(k)/\hbar$ and the BZ covariance $\operatorname{Cov}_k[f_1,f_2]=\langle f_1f_2\rangle_{\rm BZ}-\langle f_1\rangle_{\rm BZ}\langle f_2\rangle_{\rm BZ}$, and since $\hat V$, $G^{kk}_n$, and the first two terms of Eq.~\eqref{mt:eq:Aevolved} are $T$-periodic, Eqs.~\eqref{mt:eq:VarSplit} and \eqref{mt:eq:Aevolved} give $\operatorname{Var}_n[x](t+mT)=G^{kk}_n(t)+\operatorname{Var}_k\big[\mathcal A_n(k,t)+mT\,v_n(k)\big]$, i.e.,
\begin{align}
    \operatorname{Var}_n[x](t+mT)=\operatorname{Var}_n[x](t)
    +2mT\operatorname{Cov}_k\big[\mathcal A_n(k,t),v_n(k)\big]
    +m^2T^2\operatorname{Var}_k[v_n] ,
    \label{sm:eq:Var-cycles}
\end{align}
a quadratic polynomial in the cycle number $m$. The $m^2$ term describes ballistic spreading. For a smooth quasienergy band periodic in $k$, $\langle v_n\rangle_{\rm BZ}=0$, so that the $m^2$ coefficient $T^2\langle(\partial_k\varepsilon_n)^2\rangle_{\rm BZ}/\hbar^2$ vanishes only for flat bands, and the Wannier center $\langle\mathcal A_n\rangle_{\rm BZ}$ is $T$-periodic. The linear term has no definite sign, depends on the initial Wannier gauge, and vanishes, e.g., at $t=t_0$ for a Wannier function maximally localized at $t_0$. The integrated metric remains $T$-periodic even when the total spread grows. At a fixed drive phase, measurements at three cycle numbers determine the coefficients of Eq.~\eqref{sm:eq:Var-cycles}; the constant term, $\operatorname{Var}_n[x](t)=G^{kk}_n(t)+\operatorname{Var}_k[\mathcal A_n(k,t)]$, still contains the Berry-connection variance discussed below.

\textit{Gauge-dependent contribution.---} The second term of Eq.~\eqref{mt:eq:VarSplit} is the weight that $\hat X$ transfers from $\ket{W_n(t)}$ to its translates $\ket{W_{n,R}(t)}$ by $R$ unit cells. Indeed, $\mel{W_{n,R}(t)}{\hat X}{W_n(t)}=\langle e^{ikRV_{\rm uc}}\mathcal A_n(k,t)\rangle_{\rm BZ}$ are the Fourier coefficients of the Berry connection, so that Parseval's theorem gives
\begin{align}
    \operatorname{Var}_k[\mathcal A_n(k,t)]=\sum_{R\neq0}\big|\mel{W_{n,R}(t)}{\hat X}{W_n(t)}\big|^2 ,
    \label{sm:eq:VarA-real}
\end{align}
and subtracting it from the measured variance isolates $G^{kk}_n(t)$. A gauge transformation $\ket{\varphi_n(k,t_0)}\to e^{i\theta(k)}\ket{\varphi_n(k,t_0)}$ shifts $\mathcal A_n(k,t)\to\mathcal A_n(k,t)-\partial_k\theta(k)$ at all times. The maximally localized gauge makes $\mathcal A_n(k,t_0)$ $k$-independent, so that Eq.~\eqref{sm:eq:VarA-real} vanishes at $t_0$ and $\operatorname{Var}_n[x](t_0)=G^{kk}_n(t_0)$~\cite{marzari1997}. Since the gauge is fixed once the Wannier function is prepared, the remaining terms of Eq.~\eqref{mt:eq:Aevolved} cannot in general be removed at later times: the band-diagonal micromotion connection reshapes the Wannier function periodically, and the group velocity spreads it.

\textit{Band subspaces.---} For an isolated subspace $S$, let $\ket{W_j(t)}$, $j=1,\ldots,N_S$, be Wannier functions of one unit cell which, together with their translates $\ket{W_{j,R}(t)}$, span the evolved subspace with real-space projector $\cP_S(t)$. Inserting $\mathbbm{1}=\cP_S+(\mathbbm{1}-\cP_S)$ between the two position operators in $\mel{W_j}{\hat X^2}{W_j}$ gives~\cite{marzari1997}
\begin{align}
    \sum_{j=1}^{N_S}\operatorname{Var}_j[x](t)
    =\sum_{j}\mel{W_j}{\hat X\big(\mathbbm{1}-\cP_S\big)\hat X}{W_j}
    +\sum_{j}\sum_{(j',R)\neq(j,0)}\big|\mel{W_{j',R}}{\hat X}{W_j}\big|^2
    \geq G^{kk}_S(t,t_0) .
    \label{sm:eq:VarS-bound}
\end{align}
The first term is the trace per unit cell of $\cP_S\hat X(\mathbbm{1}-\cP_S)\hat X\cP_S$. In the Bloch representation, where $(\mathbbm{1}-\cP_S)\hat X\cP_S$ acts as $i(\mathbbm{1}-\cP_S)(\partial_k\cP_S)\cP_S$, it equals
\begin{align}
    \big\langle\Tr\big[\cP_S\,(\partial_k\cP_S)\,(\mathbbm{1}-\cP_S)\,(\partial_k\cP_S)\big]\big\rangle_{\rm BZ}
    =\big\langle\tfrac12\Tr\big[(\partial_k\cP_S)^2\big]\big\rangle_{\rm BZ}=G^{kk}_S(t,t_0)
\end{align}
and is independent of the choice of Wannier functions. The second term is nonnegative and, in one dimension, vanishes for the eigenfunctions of $\cP_S\hat X\cP_S$, which are the maximally localized Wannier functions~\cite{marzari1997}. A basis maximally localized at $t_0$ thus saturates Eq.~\eqref{sm:eq:VarS-bound} at $t_0$, whereas the second term is generally nonzero at later times.

\section{Flat bands, compact localized states, and their spatial variance}
\label{sm:CLS}

At $\G=1$ and $\delta=\hbar\w/2$, the A-site Nambu doublet $(\hat e_1,\hat e_3)$ has no on-site terms in Eq.~\eqref{mt:eq:bloch_ham}, and its couplings to site B combine into a rotation by the angle $k$, so that $\hh^\pm_k=e^{-ik\hat{\mathcal K}}\hh^\pm_0e^{ik\hat{\mathcal K}}$ with $\hat{\mathcal K}=i(\hat e_3\hat e_1^\top-\hat e_1\hat e_3^\top)$. The covariance carries over to products and functions of $\hh^\pm_k$, hence to $\hat U$, $\hat H_F$, and $\hat V$, so that the quasienergies are $k$-independent. The suitable gauge of the main text is $\ket{\varphi_n(k,t_0)}=e^{-ik(\hat{\mathcal K}-1)}\ket{\varphi_n(0,t_0)}$, in which $\hat X$ acts as $\hat{\mathcal K}-1$, whose eigenvalues $-2$, $-1$, $0$ give the three-site support.

The covariance of $\hat V$ gives $\hat\xi^k_V(k,t,t_0)=e^{-ik\hat{\mathcal K}}\big[\hat V_0^\dagger\hat{\mathcal K}\hat V_0-\hat{\mathcal K}\big]e^{ik\hat{\mathcal K}}$ with $\hat V_0\equiv\hat V(0,t,t_0)$, so that in this gauge the three terms of Eq.~\eqref{mt:eq:Aevolved} are $\mel{\varphi_n}{(\hat{\mathcal K}-1)}{\varphi_n}$, $\mel{\varphi_n}{\hat V_0^\dagger\hat{\mathcal K}\hat V_0-\hat{\mathcal K}}{\varphi_n}$, and zero, all evaluated at $k=0$ and thus $k$-independent. Hence $\operatorname{Var}_k[\mathcal A_n]=0$ at all times, and Eq.~\eqref{mt:eq:VarSplit} reduces to Eq.~\eqref{mt:eq:CLS-metric}. Explicitly, with $\ket{\chi_n}\equiv\ket{\chi_n(0,t,t_0)}$,
\begin{align}
    G^{kk}_n(t,t_0)=\mathrm{Var}_n[x](t)
    =\mel{\chi_n}{\hat{\mathcal K}^2}{\chi_n}-\mel{\chi_n}{\hat{\mathcal K}}{\chi_n}^2 ,
    \label{sm:eq:VarK}
\end{align}
the variance of the internal position $\hat{\mathcal K}$ in the evolved Floquet state at $k=0$. This holds for all four bands on the flat-band line $\G=1$, $\delta=\hbar\w/2$, whenever the bands remain isolated.

\textit{Invariant plane.---} With $w_1(k)=(\sin k,0,-\cos k,0)^\top=e^{-ik\hat{\mathcal K}}w_1(0)$ and $w_2=(0,0,0,1)^\top$, one finds $\hh^\pm_kw_1=w_2$ and $\hh^\pm_kw_2=w_1+(\tfrac{\hbar\w}{2}\pm\delta)w_2$ on the flat-band line, so the plane spanned by $w_1$ and $w_2$ is invariant. Within it, a state $a\,w_1(k)+b\,w_2$ evolves with $k$-independent coefficients according to the reduced Hamiltonians
\begin{align}
    i\hbar\,\partial_t\mqty(a\\b)=h^\pm\mqty(a\\b),
    \qquad
    h^\pm=\mqty(0&1\\1&\tfrac{\hbar\w}{2}\pm\delta) ,
    \label{sm:eq:hred}
\end{align}
with $h^+$ on $0<t<\tfrac T4$ and $\tfrac{3T}{4}<t<T$, and $h^-$ on $\tfrac T4<t<\tfrac{3T}{4}$.

\textit{Explicit eigenvectors and coefficients.---} At $\G=\delta=\tfrac{\hbar\w}{2}=1$, the reduced Hamiltonians of Eq.~\eqref{sm:eq:hred} are $h^+=\mathbbm{1}+\s_x-\s_z$ and $h^-=\s_x$, with Pauli matrices $\s_i$ acting in the plane, and $T=\pi\hbar$. Diagonalizing the reduced one-period evolution operator $e^{-ih^+T/4\hbar}\,e^{-ih^-T/2\hbar}\,e^{-ih^+T/4\hbar}$ gives the two negative-quasienergy flat-band Floquet states $\ket{\varphi_n(k)}=(c_n\,w_1(k)+w_2)/\sqrt{1+c_n^2}$, in a gauge that differs from the suitable gauge above by the phase $e^{-ik}$, with
\begin{align}
    c_{1,2}=-1+\dfrac{2\pm\sqrt{3+\cos(\sqrt{2}\,\pi)}}{1+\cos(\pi/\sqrt{2})},
    \label{sm:eq:c}
\end{align}
where $+$ and $-$ correspond to bands 1 and 2, labeled by increasing quasienergy. The coefficients are real, and orthogonality of the two states gives $c_1c_2=-1$. The micromotion only rotates the states within the plane, $\ket{\chi_n(k,t,t_0)}=a_n(t)\,w_1(k)+b_n(t)\,w_2$, where
\begin{align}
    \mqty(a_n(t)\\b_n(t))=e^{i\varepsilon_n(t-t_0)/\hbar}\,u(t)\,\frac{1}{\sqrt{1+c_n^2}}\mqty(c_n\\1) ,
    \label{sm:eq:abt}
\end{align}
and $u(t)$ is the time-ordered product of the reduced propagators $e^{-ih^\pm t/\hbar}$ from $t_0=0$ to $t$.

\textit{Compact localized states.---} Following Ref.~\cite{rhim2019classification}, superposing these Bloch states with envelope $e^{ik}$, i.e., the states of the suitable gauge with a uniform envelope, yields the CLS of band $n$ on three consecutive sites, with Bogoliubov operator
\begin{align}
    -\tfrac{a_n}{2}\big(if^\dagger_{2\ell-2}+f_{2\ell-2}\big)
    +b_n\,f_{2\ell-1}+\tfrac{a_n}{2}\big(if^\dagger_{2\ell}-f_{2\ell}\big),
    \label{sm:eq:CLS-op}
\end{align}
where $a_n=a_n(t)$ and $b_n=b_n(t)$. Its particle and hole components $u^n_m(t)$ and $v^n_m(t)$ on site $m$, the coefficients of $f^\dagger_m$ and $f_m$, define the BdG weight $\rho^n_m(t)=|u^n_m(t)|^2+|v^n_m(t)|^2$.

\section{Detection of the compact localized states}
\label{sm:experiment}

On the flat-band line $\G=1$, $\delta=\hbar\w/2$, each rotating-frame bond is an Ising coupling along the anisotropy axis and the A-site field $\delta-\hbar\w/2$ vanishes (Sec.~\ref{sm:chiralspinchain}). The A-spin components along this axis are therefore conserved, and the chain decouples into three-site clusters. The cluster modes associated with $w_1(k)$ and $w_2$ of Sec.~\ref{sm:CLS} through the envelope $e^{ik}$,
\begin{align}
    d^\dagger_{\ell,1}=\tfrac12\big(if^\dagger_{2\ell}-f_{2\ell}\big)-\tfrac12\big(if^\dagger_{2\ell-2}+f_{2\ell-2}\big),
    \qquad
    d^\dagger_{\ell,2}=f_{2\ell-1} ,
    \label{sm:eq:dmodes}
\end{align}
turn Eq.~\eqref{sm:eq:CLS-op} into $a_nd^\dagger_{\ell,1}+b_nd^\dagger_{\ell,2}$, whose coefficients evolve under Eq.~\eqref{sm:eq:hred}. A cluster hosts a single excitation exactly when its two A spins are parallel along the anisotropy axis. Since $d^\dagger_{\ell,2}d_{\ell,2}=\tfrac12(1+\s^z_{2\ell-1})$, $|b_n|^2$ and $|a_n|^2$ are then the probabilities to find the central spin up and down, so that
\begin{align}
    G^{kk}_n(t)=\mathrm{Var}_n[x](t)=|a_n(t)|^2=\frac{1-\langle\s^z_{2\ell-1}\rangle(t)}{2} .
    \label{sm:eq:G-sz}
\end{align}
The CLS of band $n$ can thus be prepared locally using a three-spin product state: at $t=0$, where the laboratory and rotating frames coincide, the two outer A spins are polarized parallel to each other along $(\mathbf e_x-\mathbf e_y)/\sqrt2$, and the central B spin is prepared in the Floquet eigenstate of its conditional two-level evolution that corresponds to band $n$ [Eq.~\eqref{sm:eq:c}]. Its magnetization over one drive period then yields $G^{kk}_n(t)$ in either frame, since $\hat R(t)$ commutes with $\s^z_m$.

The full weight distribution $\rho^n_m(t)$ follows from the normal and anomalous correlators $\langle f^\dagger_mf_{m'}\rangle$ and $\langle f_mf_{m'}\rangle$ on the three-site support. These are local spin correlators with at most one intermediate Jordan--Wigner string factor, e.g., $f^\dagger_mf_{m+2}=\s^-_m\s^z_{m+1}\s^+_{m+2}$ and $f_{m+2}f_m=\s^+_m\s^z_{m+1}\s^+_{m+2}$, and require no external phase reference. With $\mathcal{C}_{\mu\nu}=\langle\Psi_\mu\Psi^\dagger_\nu\rangle$ for $\Psi=(f_{2\ell-2},f_{2\ell-1},f_{2\ell},f^\dagger_{2\ell-2},f^\dagger_{2\ell-1},f^\dagger_{2\ell})^\top$, and $\mathcal W=(\hat w_1,\hat w_2)$ the vectors of the cluster modes~\eqref{sm:eq:dmodes}, $\hat w_1=\tfrac12(-i,0,i,-1,0,-1)^\top$ and $\hat w_2=(0,0,0,0,1,0)^\top$ in the basis of $\Psi^\dagger$, the matrix $\mathbbm{1}_2-\mathcal W^\dagger\mathcal C\,\mathcal W$ is the rank-one projector onto $(a_n,b_n)^\top$. Its diagonalization yields $a_n(t)$ and $b_n(t)$ up to a common phase and, through Eq.~\eqref{sm:eq:CLS-op}, the weights $\rho^n_m(t)$, which give $\mathrm{Var}_n[x](t)=G^{kk}_n(t)$ (End Matter~\ref{mt:em:CLS}).

\section{Numerical calculation}
\label{sm:numcal}

We set $\hbar=1$ and measure energies in units of the exchange coupling of Eq.~\eqref{mt:eq:spinHam0}. All figures use the same numerical parameters: $N_k=501$ equally spaced momenta in $[-\tfrac\pi2,\tfrac\pi2)$, centered differences between neighboring momenta, $\delta k=\pi/N_k$, and $N_t=240$ time steps per period, $\Delta t=T/N_t$, with the exact step propagators $e^{-i\hh^\pm\Delta t/\hbar}$. All real-space calculations use 100 sites (50 unit cells) at positions $x=1,2,\ldots,100$, with $B_\ell$ on site $2\ell-1$ and $A_\ell$ on site $2\ell$.

\textit{Momentum-space quantities [Figs.~\ref{mt:fig:4B_results_strobo}(b,c) and \ref{mt:fig:4B_results_micro}(a,b)].---} We evaluate $\hat U(k,t,t_0)$ from the step propagators of Eq.~\eqref{mt:eq:bloch_ham}. Diagonalizing $\hat U(k,t_0+T,t_0)$ yields the quasienergies in the zone $(-\hbar\w/2,\hbar\w/2]$ and the projectors $\hat P_n(k,t_0)$, hence $\hat H_F(k,t_0)$ and, from Eq.~\eqref{mt:eqn:micromotionOperatorDefinition}, $\hat V(k,t,t_0)$ and the evolved projectors of Eq.~\eqref{mt:eqn:timeFloquetProjector}. We compute $g^{kk}_S=\tfrac12\Tr[(\partial_k\cP_S)^2]$ from centered differences of the gauge-invariant projector, and $G^{kk}_S$ as its average over the grid. We evaluate the projectors and $\hat V$ at $k\pm\delta k$ from the evolution generated by Eq.~\eqref{mt:eq:bloch_ham} at each shifted momentum, even when one of these points lies outside the BZ. The micromotion connection $\hat\xi^k_V=i\hat V^\dagger\partial_k\hat V$ follows from centered differences of $\hat V$, which is smooth in $k$ for an open $\pi$-gap, and gives $M^{kk}_S$. The bounds of Fig.~\ref{mt:fig:4B_results_strobo}(c) are Eq.~\eqref{sm:eq:envelope} with $t_0=0$ and Eq.~\eqref{mt:eq:shared} with $t_s=T/2$, the latter with the micromotion operator $\hat V(k,t,T/2)=\hat V(k,t,0)\hat V^\dagger(k,T/2,0)$ referenced at $T/2$. The windings $\nu_0$ and $\nu_\pi$, and hence $\nu_s=\nu_0\pm\nu_\pi$ in Eq.~\eqref{mt:eq:shared}, are those of the determinants of the off-diagonal and diagonal blocks of the half-period evolution operator $U_1$ in the chiral basis~\cite{asboth2014chiral}.

\textit{Time-evolved Wannier functions [Figs.~\ref{mt:fig:4B_results_strobo}(d) and \ref{mt:fig:4B_results_micro}(d)].---} We construct maximally localized Wannier functions (MLWFs) for an isolated subspace $S$ of $N_S$ Floquet bands. Here $S$ is either the chiral subspace ($N_S=2$) or a single band ($N_S=1$). In one dimension, the MLWFs of $S$ are the eigenvectors of the projected position operator~\cite{marzari1997},
\begin{align}
    \hat P_S\hat X\hat P_S\ket{W_j}=\bar x_j\ket{W_j} ,
    \label{sm:eq:SPXP}
\end{align}
where $\hat X$ assigns both Nambu components of site $m$ its position $x_m=m$, and $\bar x_j$ is the center of $\ket{W_j}$. We take
\begin{align}
    \hat P_S=\sum_{\varepsilon_\lambda\in\mathcal I_S}\ket{\phi_\lambda}\bra{\phi_\lambda} ,
    \label{sm:eq:SPS}
\end{align}
the spectral projector onto the quasienergy range $\mathcal I_S$ of $S$ of the one-period evolution operator of the ring, i.e., the chain closed with periodic boundary conditions for the fermions, $\hat U(T,0)\ket{\phi_\lambda}=e^{-i\varepsilon_\lambda T/\hbar}\ket{\phi_\lambda}$. The ring has no edge modes, and $\hat P_S$ is the Fourier transform of the Bloch projector $\hat P_S(k,0)$. Among the eigenvectors of Eq.~\eqref{sm:eq:SPXP}, we keep the $N_S$ whose centers lie in one unit cell near the chain center, $j=1,\dots,N_S$. For $S=\mathrm{ch}$, these are Wannier functions of the subspace rather than of the individual bands. These states evolve as $\ket{W_j(t)}=\hat U(t,0)\ket{W_j}$, with $\hat U$ built from the step propagators of the open chain. Their variances follow from the BdG weights,
\begin{align}
    \mathrm{Var}_j[x](t)=\sum_m\rho^j_m(t)\big[x_m-\bar x_j(t)\big]^2 ,\quad
    \bar x_j(t)=\sum_m\rho^j_m(t)\,x_m ,
    \label{sm:eq:SVarj}
\end{align}
with $\rho^j_m=|u^j_m|^2+|v^j_m|^2$, where $u^j_m$ and $v^j_m$ are the particle and hole components of $\ket{W_j(t)}$ on site $m$, normalized to $\sum_m\rho^j_m=1$. The total spread of the unit cell,
\begin{align}
    \mathrm{Var}_S(t)=\sum_{j=1}^{N_S}\mathrm{Var}_j[x](t)\geq G^{kk}_S(t,0) ,
    \label{sm:eq:SVarS}
\end{align}
equals $G^{kk}_S(0)$ at $t=0$ [Eq.~\eqref{sm:eq:VarS-bound}], and we denote $\mathrm{Var}_{\rm ch}\equiv\mathrm{Var}_S$ for $S={\rm ch}$.

\textit{Compact localized states [Figs.~\ref{mt:fig:4B_results_micro}(b,c)].---} At the flat-band point, the CLS of band $n$ is the real-space vector $a_n(0)\hat w_1+b_n(0)\hat w_2$ of Eqs.~\eqref{sm:eq:CLS-op} and \eqref{sm:eq:dmodes}, placed on three sites near the chain center and evolved like the Wannier functions. Its site-resolved weight $\rho^n_m(t)$ and variance follow from Eq.~\eqref{sm:eq:SVarj}.

\bibliography{refs_updated}